\documentclass[10pt,twocolumn,superscriptaddress,nofootinbib,pre]{revtex4-1}

\usepackage{amsmath}
\usepackage{amssymb}
\usepackage{graphicx}
\usepackage{color}
\usepackage{tikz}
\usepackage{url}
\usepackage{xr}
\usepackage{enumitem}
\usepackage{mathtools}
\usepackage{bm}
\usepackage{float}
\usepackage{enumitem}
\usepackage{fancyvrb}
\usepackage{nameref,hyperref}
\setlist[enumerate,1]{start=1}

\newcommand{\ct}{(i,j)_t}
\newcommand{\Dt}{{\Delta t}}
\renewcommand{\S}{{\mathcal S}}
\newcommand{\I}{{\mathcal I}}
\newcommand{\R}{{\mathcal R}}

\renewcommand{\l}[1]{\lambda_{#1}}
\newcommand{\X}[1]{\mathcal{X}_{#1}}

\newcommand{\beq}{\begin{equation}}
\newcommand{\eeq}{\end{equation}}

\newcommand{\kIt}{{k_{\I}(t)}}
\newcommand{\Oa}{{\Omega}}

\newcommand{\Lat}{{\Lambda(t)}}
\newcommand{\La}[1]{{\Lambda(#1)}}
\newcommand{\es}{\enspace}
\newcommand{\F}[1]{{\mathcal{F}_t^{(m)}}}
\newcommand{\Ft}{{\mathcal{F}_t}}
\renewcommand{\t}[1]{{t^{(#1)}}}
\newcommand{\ts}{{t^{*}}}
\newcommand{\tss}{{t^{**}}}
\newcommand{\ns}{{n^{*}}}
\newcommand{\nss}{{n^{**}}}

\newcommand{\Lt}{{\mathbb{L}(t;\ts)}}

\renewcommand{\L}{\mathbb{L}}

\newcommand{\Lalgo}{{\Lambda\Dt}}
\renewcommand{\Im}{{{\rm I}_m}}
\newcommand{\kt}{{\overline{k(t)}}}
\newcommand{\taum}{{\langle\tau^{(m)}\rangle}}
\newcommand{\Exp}{{\rm Exp}}
\newcommand{\Ord}[1]{\mathcal{O}\left(#1\right)}
\newcommand{\Trs}{\Theta_{\rm RS}}
\newcommand{\Ttga}{\Theta_{\rm TGA}}
\newcommand{\nsimu}{{n_{\rm simu}}}

\newcommand{\Emean}{\overline{E(t)}}
\newcommand{\Mmean}{\overline{M(t)}}
\newcommand{\Qmean}{\overline{Q(t)}}
\newcommand{\Msi}{\overline{M_{\rm S\to I}(t)}}
\newcommand{\Imean}{\overline{I(t)}}

\begin{document}

\title{Temporal Gillespie algorithm: Fast simulation of contagion processes on time-varying networks}
\date{\today}
\author{Christian L. Vestergaard}
\email{cvestergaard@gmail.com}
\affiliation{Aix Marseille Universit\'e, Universit\'e de Toulon, CNRS, CPT, UMR 7332, 13288 Marseille, France}
\author{Mathieu G\'enois}
\affiliation{Aix Marseille Universit\'e, Universit\'e de Toulon, CNRS, CPT, UMR 7332, 13288 Marseille, France}

\begin{abstract}
Stochastic simulations are one of the cornerstones of the analysis of dynamical processes on complex networks, and are often the only accessible way to explore their behavior.
The development of fast algorithms is paramount to allow large-scale simulations.
The Gillespie algorithm can be used for fast simulation of stochastic processes, and variants of it have been applied to simulate dynamical processes on static networks.
However, its adaptation to temporal networks remains non-trivial.
We here present a temporal Gillespie algorithm that solves this problem.
Our method is applicable to general Poisson (constant-rate) processes on temporal networks, stochastically exact, and up to multiple orders of magnitude faster than traditional simulation schemes based on rejection sampling.
We also show how it can be extended to simulate non-Markovian processes.
The algorithm is easily applicable in practice, and as an illustration we detail how to simulate both Poissonian and non-Markovian models of epidemic spreading. Namely, we provide pseudocode and its implementation in C++ for simulating the paradigmatic Susceptible-Infected-Susceptible and Susceptible-Infected-Recovered models and a Susceptible-Infected-Recovered model with non-constant recovery rates.
For empirical networks, the temporal Gillespie algorithm is here typically from 10 to 100 times faster than rejection sampling.
\end{abstract}

\maketitle

% Please keep the Author Summary between 150 and 200 words
% Use first person. PLOS ONE authors please skip this step.
% Author Summary not valid for PLOS ONE submissions.
\section*{Author Summary}
When studying how e.g. diseases spread in a population, intermittent contacts taking place between individuals{---}through which the infection spreads{---}are best described by a time-varying network. This object captures both their complex structure and dynamics, which crucially affect spreading in the population. The dynamical process in question is then usually studied by simulating it on the time-varying network representing the population. Such simulations are usually time-consuming, especially when they require exploration of different parameter values.
We here show how to adapt an algorithm originally proposed in 1976 to simulate chemical reactions---the Gillespie algorithm{---}to speed up such simulations. Instead of checking at each time{-}step if each possible reaction takes place, as traditional rejection sampling algorithms do, the Gillespie algorithm determines what reaction takes place next and at what time. This offers a substantial speed gain by doing away with the many rejected trials of the traditional methods, with the added benefit of giving stochastically exact results. In practice this new temporal Gillespie algorithm is tens {to hundreds} of times faster than the current state-of-the-art, opening up for thorough characterization of spreading phenomena and fast large-scale applications such as the simulation of city- or world-wide epidemics.

\section{Introduction}
Networks have emerged as a natural description of complex systems and their dynamics~\cite{Barrat2008}, notably in the case of spreading phenomena, such as social contagion, rumor and information spreading, or epidemics~\cite{Barrat2008,Balcan2009,Pastor-Satorras2014}.
The dynamics of contagion processes occurring on a network are usually complex, and analytical results are attainable only in special cases~\cite{Pastor-Satorras2014,Ferreira2012}.
Furthermore, such results almost systematically involve approximations~\cite{Pastor-Satorras2014,Ferreira2012}.
Numerical studies based on stochastic simulations are therefore necessary, both to verify analytical approximations, and to study the majority of cases for which no analytical results exist.
The development of fast algorithms is thus important for the characterization of contagion phenomena, and for large-scale applications  such as simulations of world-wide epidemics~\cite{Balcan2009,Tizzoni2012}.

The Doob-Gillespie algorithm~\cite{Doob1942,Doob1945,Kendall1950,Bartlett1953,Gillespie1976,Gillespie1977} (also known as Gillespie's {\sl Stochastic Simulation Algorithm---SSA} or {\sl Gillespie's direct method}), originally proposed by David Kendall in 1950 for simulating birth-death processes and made popular by Daniel Gillespie in 1976 for the simulation of coupled chemical reactions, offers an elegant way to speed up such simulations by doing away with the many rejected trials of traditional Monte Carlo methods. %based on rejection sampling.
Instead of checking at each time-step if each possible reaction takes place, as rejection sampling algorithms do, the Gillespie algorithm draws directly the time elapsed until the next reaction takes place and what reaction takes place at that time.
It is readily adapted to the simulation of Poisson processes on static networks~\cite{Huerta2002,Dangerfield2009,Hladish2012,Holme2014,Zschaler2012} and can be generalized to non-Markovian processes~\cite{Boguna2014}.

Systems in which spreading processes take place, e.g., social, technological, infrastructural, or ecological systems, are not static though. Individuals create and break contacts at time-scales comparable to the time-scales of such processes~\cite{Onnela2007,Rybski2009,Cattuto2010}, and
the dynamics of the networks themselves thus profoundly affect dynamical processes taking place on top of them~\cite{Vazquez2007,Miritello2011a,Karsai2011b,Panisson2012,Gauvin2013,Holme2014a,Karsai2014}.
%To get the full picture of the phenomenology, o
{This means that one} needs to take the network's dynamics into account, e.g., by representing it as a time-varying network (also known as a time-varying graph, temporal network, or dynamical network)~\cite{Holme2012}.
The dynamical nature of time-varying networks makes the adaptation of the Gillespie algorithm to such systems non-trivial.

The main difficulty in adapting the Gillespie algorithm to time-varying networks is taking into account the variation of the set of possible transitions and of their rates at each time step.
We show that by normalizing time by the instantaneous cumulative transition rate, we can construct a temporal Gillespie algorithm that is applicable to Poisson (constant rate) processes on time-varying networks.
%The method thus constructed is easily applicable in practice.
We give pseudocode and C++ implementations for its application to simulate the paradigmatic Susceptible-Infected-Susceptible (SIS) and Susceptible-Infected-Recovered (SIR) models of epidemic spreading, for both homogeneous and heterogeneous~\cite{Cai2013} populations.
%The resulting algorithm is up to orders of magnitude faster.
%The accuracy of the temporal Gillespie algorithms is verified numerically by comparison with a classical rejection sampling algorithm.
We verify the accuracy of the temporal Gillespie algorithm numerically by comparison with a classical rejection sampling algorithm, and we show that it is up to {$\sim 500$ times} faster {for the processes and the parameter ranges investigated here}.

While Poissonian models are of widespread use, real contagion phenomena show memory effects, i.e., they are non-Markovian.
Notably, for realistic infectious diseases, the rate at which an infected individual recovers is not constant over time~\cite{Ferguson2006,Lloyd2001}. 
Social contagion may also show memory effects, e.g., one may be more (or less) prone to adopt an idea the more times one has been exposed to it.
To treat this larger class of models, we show how the temporal Gillespie algorithm can be extended to non-Markovian processes.
We give in particular an algorithm for simulating SIR epidemic models with non-constant recovery rates.

% Results and Discussion can be combined.
\section{Results}
The following subsections present the main results of the article.
Section~\ref{sec:processes} defines the stochastic processes which can be simulated using the temporal Gillespie algorithm, and describes the class of compartmental models for contagion phenomena on networks{---}the class we will use in examples throughout this paper.
Section~\ref{sec:rejectionSampling} gives a quick overview of the traditional rejection sampling algorithms.
Section~\ref{sec:staticGillespie} outlines a derivation of the static Gillespie algorithm.
Section~\ref{sec:temporalGillespie} derives the temporal Gillespie algorithm for Poisson (constant-rate) processes.
In Section~\ref{sec:comparison} we validate the temporal Gillespie algorithm through numerical comparison with a rejection sampling algorithm; we also compare their speeds for simulating SIR and SIS processes on {both} synthetic {and empirical} time-varying networks.
Section~\ref{sec:nonMarkovian} shows how the temporal Gillespie algorithm can be extended to simulate non-Markovian processes; the approach is verified numerically and the speed of the non-Markovian temporal Gillespie algorithm is compared to rejection sampling.

Tables listing the notation used in the manuscript, {details on how Monte Carlo simulations were performed, and pseudocode for application of the temporal Gillespie algorithm} are given in the {Methods} section.

\subsection{Stochastic processes on time-varying networks}
\label{sec:processes}
We define in this section the type of stochastic processes {for} which the temporal Gillespie algorithm can be applied.
At the time of writing, the main domain of application of the algorithm is the class of compartmental models for contagion processes on time-varying networks, which we introduce below.
For definiteness, algorithms detailing the application of the temporal Gillespie algorithm will concern this class of stochastic processes.

In general, we consider a system whose dynamics can be described by a set of stochastic transition events.
We assume that the system can be modeled at any point in time by a set, $\Oa(t)$, of $M(t)$ independent stochastic processes $m$, which we term {\sl transition processes}; the rate at which the transition $m$ takes place is denoted $\l{m}$.
{The set $\Oa(t)$ thus defines the possible transition events at time $t$ and}
in general changes over time,
{depending on both external factors and the evolution of the system itself;
the number of possible transitions, $M(t)$, thus also generally changes over time, while}
$\l{m}$ may or may not vary over time.
For the classic ``static'' Gillespie algorithm to be applicable, $\Oa(t)$ {is allowed to} change only when a transition (or chemical reaction in the context of Gillespie's original article) takes place.
For processes taking place on time-varying networks, the medium of the process---the network---also changes with time.
As these changes may allow or forbid transitions, $\Oa(t)$ is not only modified by every reaction, but also by every change in the network.
This is the domain of the temporal Gillespie algorithm, which only requires that the number of points in which $\Oa(t)$ changes be finite over a finite time-interval~{\cite{footnote1}}.

The assumption that the transition processes are independent is essential to the validity of the Gillespie algorithm, as it allows the calculation of the distribution of waiting times between consecutive transitions.
This assumption is not overly restrictive, as it only requires a transition process to be independent of the evolution of the other simultaneous transition processes.
A transition process may depend on all earlier transitions, and the current and past states of all nodes.
As such, Gillespie algorithms can notably be applied to models of cooperative infections and other non-linear processes such as threshold models~\cite{Boguna2014}, and {has even been applied to model} the dynamics of ant battles~\cite{Martelloni2015}.

\begin{figure}
%\begin{adjustwidth}{-2.25in}{0in} % Comment out/remove adjustwidth environment if table fits in text column.
   \centering
   \includegraphics[width=0.48\textwidth]{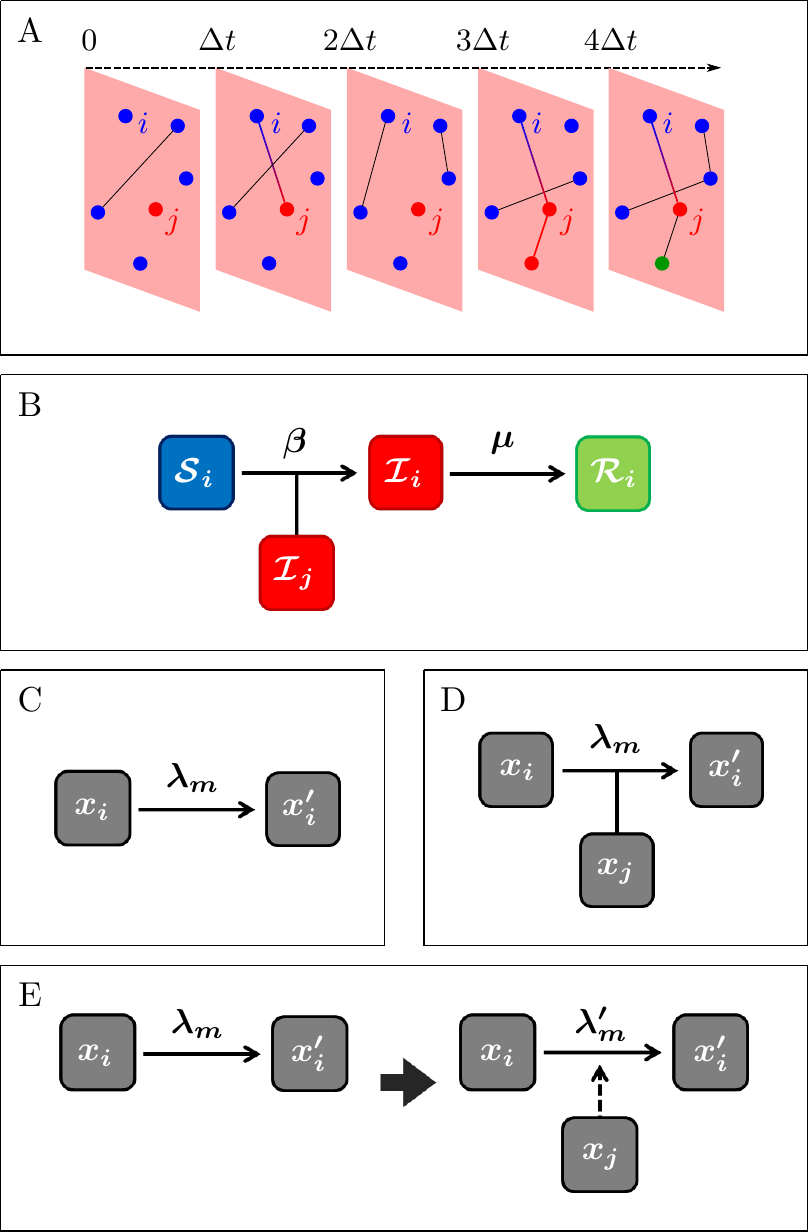}
   \caption{{\bf Schematic representation of a compartmental contagion process on a network.}
   (A) {Illustration of a contagion process evolving on a time-varying network. Nodes' colors correspond to their current state; edges denote current contacts between nodes; edge colors correspond to: black: no contagion may take place over the edge, red: contagion takes place {during the present time-step}, and red-to-blue gradient: contagion is possible but does not take place.}
   (B) Example: reaction types in the SIR model.
   (C) Spontaneous reaction: a node $i$ may spontaneously transition from its current state $x_i$ to $x'_i$ with rate $\l{m}$.
   (D) Contact-dependent reaction: a node $i$ may transition from its current state $x_i$ to $x_i'$ with rate $\l{m}$ upon contact with the node $j$ in state $x_j$.
   (E) Mixed transition: a node $i$ may spontaneously transition from its current state $x_i$ to another state, $x_i'$ with rate $\l{m}$; contact with another node $j$, in state $x_j$, may alter the transition rate of $m$, $\l{m}\to\l{m}'$. After the contact $\ct$ ends, the transition rate may revert to $\l{m}${, remain unchanged,} or change to a third value.
   }
   \label{fig:model}
%\end{adjustwidth}
\end{figure}

\paragraph*{Compartmental models of contagion.}
\label{sec:compartmental}
In a network-based description of the population in which a contagion process takes place, an individual is modeled as a node $i$ (Fig.~\ref{fig:model}A).
A contact between two individuals taking place at time $t$ is represented by an edge $(i,j)_t$ in the graph describing the population at the instant $t$ (Fig.~\ref{fig:model}A).
In a compartmental model, each node is in a certain state, which belongs to a fixed, finite set of $q$ different states (compartments)~\cite{Pastor-Satorras2014}.
A random variable $x_i(t)\in\{\X{1},\X{2},\ldots,\X{q}\}$ specifies the state of the node $i$ at time $t$ (i.e. to which compartment it belongs).
Nodes may stochastically transition between states, governed by the set $\Oa(t)$ of transition processes.
One is usually interested in the evolution of the number of nodes in each state, which we denote $X_1,X_2,\ldots,X_q$.

As an example, we consider the classic SIR model of epidemic spreading with constant transition rates in a homogeneous population (rates are the same for all individuals) (Fig.~\ref{fig:model}B). Here nodes can be in one of three states: susceptible, infected, and recovered, $\{\S,\I,\R\}$. Two different transition types let nodes change state: (i) a node $i$ in the $\S$ state switches to the $\I$ state with rate $\kIt\beta$ (an $\S\to\I$ reaction), where $\kIt$ is the number of contacts $i$ has with nodes in the $\I$ state {at time $t$} (Fig.~\ref{fig:model}A); (ii) a node $i$ in the $\I$ state switches to the $\R$ state at rate $\mu$ (an $\I\to\R$ reaction).
%We see that the $\S\to\I$ transition is a contact-dependent reaction (type {\sl b}) and that the $\I\to\R$ transition is a spontaneous reaction (type {\sl a}).
%Figure~\ref{fig:model}E shows the evolution of the number of nodes in the $\S$, $\I$, and $\R$ states---$S$, $I$, and $R$, respectively ---for 1\,000 realizations of an SIR processes on a time-varying network.

In general, the transition processes can be divided into three types:
\renewcommand{\theenumii}{\alph{enumi}}
\begin{enumerate}[label=\alph*.]
  \item spontaneous transitions, which only depend on the current state of the node, $x_i(t)$ (Fig.~\ref{fig:model}C)---e.g. an infected node recovers spontaneously in the SIR model (Fig.~\ref{fig:model}B);
  \item contact-dependent transitions, which may take place only when the node $i$ is in contact with other nodes in a given state; it thus depends on the states {$x_j$} of the {nodes $j$ currently in contact with $i$} (Fig.~\ref{fig:model}D)---e.g. a susceptible node may be infected upon contact with an infectious node in the SIR model (Fig.~\ref{fig:model}B).
  \item mixed transitions, which take place spontaneously, but may depend on the node's past and current contacts (Fig.~\ref{fig:model}E)---e.g. in rumor spreading, an individual may learn on his own that a rumor is false (spontaneous) or may be convinced by another individual who knows the rumor is false (contact-dependent).
\end{enumerate}
This division is important for practical application of the temporal Gillespie algorithm as transition processes of type {\sl a} need only be updated after a transition has taken place, and processes of type {\sl c} need only be updated whenever a relevant contact takes place, but not at each time-step. Using this distinction is crucial to obtaining the large speed-increase that the temporal Gillespie algorithm offers over rejection sampling, as discussed below (Secs.~\ref{sec:temporalGillespie} and \ref{sec:comparison}).

\begin{figure*}
%\begin{adjustwidth}{-2.25in}{0in} % Comment out/remove adjustwidth environment if table fits in text column.
   \centering
   \includegraphics[width=.8\textwidth]{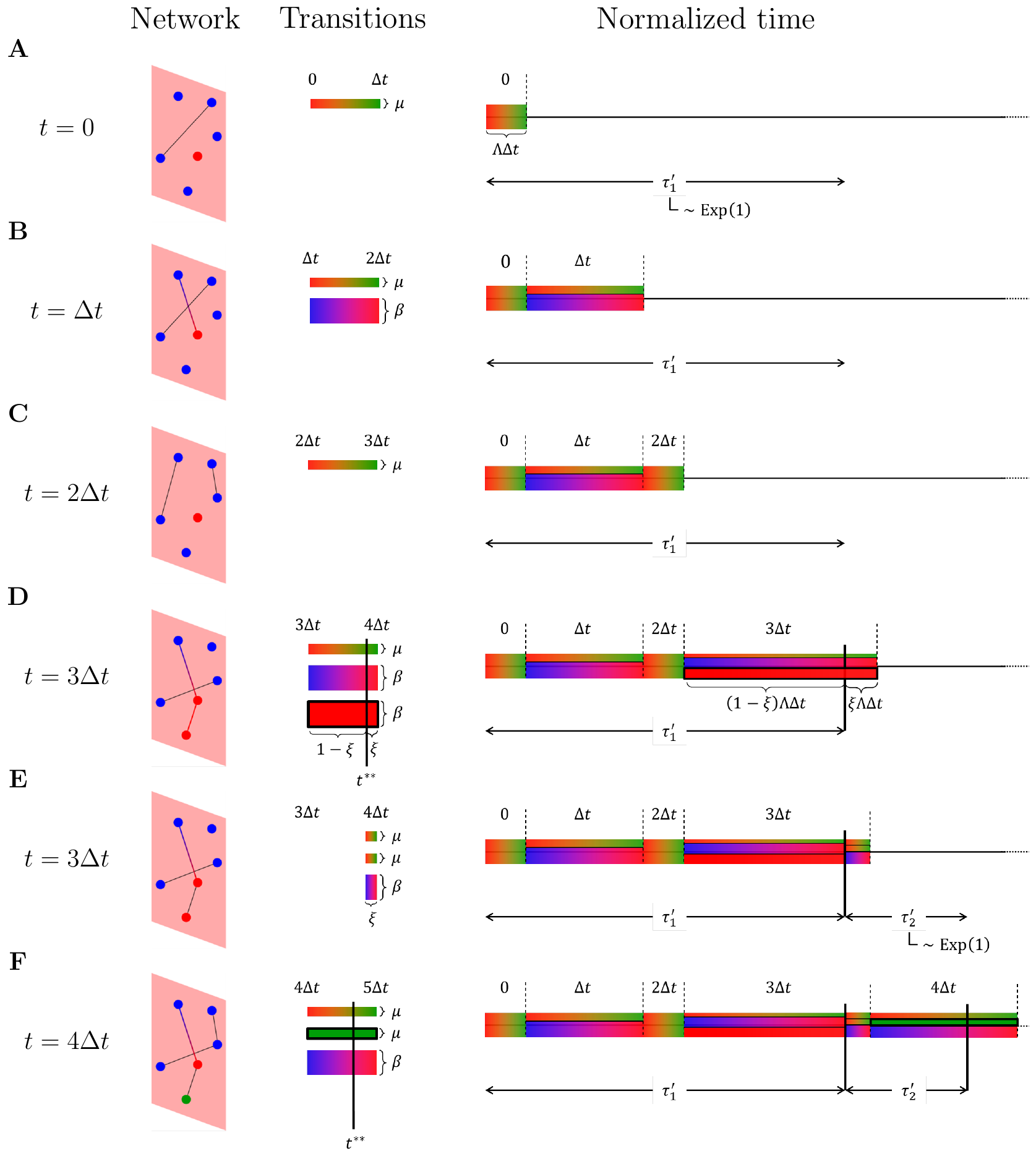}
   \caption{{{\bf Example of how the temporal Gillespie algorithm works for an SIR process on a time-varying network.}
   We consider the time-varying network of Fig~\ref{fig:model}A (Network)---time evolves along the vertical axis; a rejection sampling algorithm considers each transition process at each time-step individually (Transitions); the temporal Gillespie algorithm considers the integrated cumulative transition rate of all transition processes, $\L(t;0)$, and compares it with the random normalized waiting times $\tau_l'$ (Normalized time). A transition takes place whenever $\L(t;0)=\sum_{l=0}^q\tau_l'$ for any $q\in\mathbb{N}$.
   The temporal Gillespie algorithm works as follows.
   (A) The first normalized waiting time $\tau_1'$ is drawn from an exponential distribution with unit rate [$\tau_1'\sim\Exp(1)$] (Normalized time). From the state of the network at the first time-step, the set of possible transitions $\Oa(0)$ is found (Transitions), and from this the cumulative transition rate $\Lambda(0)$ is calculated. The integrated cumulative transition rate, $\L(\Dt;0)=\Lambda(0)\Dt$ is compared to $\tau_1'$. If, as in the present example, $\Lambda(0)\Dt<\tau_1'$ %, then $\Lambda\Dt$ is subtracted from $\tau_1'$ and
   the algorithm is advanced to the next time-step.
   (B) and (C) The set of possible transitions $\Oa(t_n)$ and the cumulative transition rate $\La{t_n}$ is updated at each following time-step $n$; if $\L(t_n;0)=\Dt\sum_{l=0}^{n-1}\La{t_l}$ is still smaller than $\tau_1'$, the algorithm is advanced to the next time-step.
   (D) During the first time-step $\nss$ where $\L(t_{\nss};0)\geq\tau_1'$, a transition takes place. The exact time of this transition, $\tss$, is given by Eq.~(\ref{eq:tss}) and the transition $m$ that takes place is chosen among the possible transitions in the given time-step with probability proportional to its transition rate $\l{m}$ [Transitions and Eq.~(\ref{eq:pi_m(t)})].
   (E) The transition changes the system (Network) and consequently the set of possible transitions, $\Oa(\tss)$, (Transitions); thus $\Oa(\tss)$ and $\La{\tss}$ must be updated, which in turn changes the remainder of $\L(t_{\nss};0)$ (Normalized time). A new normalized waiting time is then drawn, $\tau_2'\sim\Exp(1)$; if $\L(t_{\nss};0)<\tau_1'+\tau_2'$, no further transitions takes place during the time-step and the algorithm is advanced to the next time-step (note that multiple transitions may occur during the same time-step).
   (F) The above procedure is reiterated.}
   }
   \label{fig:algo}
%\end{adjustwidth}
\end{figure*}
\clearpage

\subsection{Rejection sampling for Monte Carlo simulations}
\label{sec:rejectionSampling}
A straightforward way to simulate a stochastic process is to use a rejection sampling algorithm, akin to the classical Metropolis algorithm.
Here one divides the time-axis in small time-steps $\Dt$, where $\Dt$ should be chosen sufficiently small such that this discretization does not influence the outcome of the process significantly; on time-varying networks, the choice of $\Dt$ often comes naturally as the time-resolution at which the network is measured or simulated (Fig.~{\ref{fig:model}}A).

At each time-step $t=0,\Dt, 2\Dt, \ldots$, we test whether each possible transition $m\in\Oa(t)$ takes place or not.
In practice this is done by drawing a random number $r_m$ that is uniformly distributed on $[0,1)$ for each $m$ and comparing it to $\l{m}\Dt$: if $r_m<\l{m}\Dt$ the reaction takes place, if $r_m\geq\l{m}\Dt$ nothing happens {[Fig.~\ref{fig:algo} (Transitions)]}.
(Note that one should technically compare $r_m$ to $1-\exp(\l{m}\Dt)$ to ensure that $\l{m}$ defines a proper transition rate for finite $\Dt$. However, the two procedures are equivalent in the limit $\Dt\to0$.)

From the design of the rejection sampling algorithm we see that the proportion of trials that are rejected is equal to a weighted average over $\{1-\l{m}\Dt\}_m$. Thus, since we require $\l{m}\Dt\ll1$ in order to avoid discretization errors, the vast majority of trials are rejected and the rejection sampling algorithm is computationally inefficient.

\subsection{Gillespie algorithm on static networks}
\label{sec:staticGillespie}
The Gillespie algorithm lets us perform stochastically exact Monte Carlo simulations without having to reject trials.
For Poisson processes on static networks, it works by recognizing that  the waiting time between two consecutive transitions is exponentially distributed, and that each transition happens with a probability that is proportional to its rate.

Specifically, the (survival) probability that the transition $m$ has not taken place after a time $\tau$ since the last transition event is
\beq
  S_m(\tau) = e^{-\l{m}\tau} \es.
  \label{eq:S_m:static}
\eeq
Since each transition takes place independently, the probability that no event takes place during the interval $\tau$ since the last event is
\beq
  S(\tau) = \prod_m S_m(\tau) = e^{-\Lambda\tau} \es,
\eeq
where $\Lambda=\sum_{m=1}^M \l{m}$ is the cumulative transition rate. The above result is obtained by using the fact that while $\Oa$ and $M$ do depend on $t$, they only change when an event takes place and not in-between. They can thus be treated as constant for the purpose of calculating the waiting time between events.
The distribution of the waiting times $\tau$ is then {given by the probability density} $p(\tau)=\Lambda e^{-\Lambda\tau}$, while the probability {density for the} reaction $m$ {being} the next reaction that takes place and that it takes place after exactly time $\tau$ is equal to $p_m(\tau)=\l{m}e^{-\Lambda\tau}$

The static Gillespie algorithm thus consists in drawing the waiting time $\tau\sim\Exp(\Lambda)$ until the next transition and then drawing which transition $m$ takes place with probability $\pi_m=\l{m}/\Lambda$. [Here $\tau\sim\Exp(\Lambda)$ is short for: $\tau$ is exponentially distributed with {rate $\Lambda$}.]

\subsection{Temporal Gillespie algorithm}
\label{sec:temporalGillespie}
For processes taking place on time-varying networks however{, the set of transition process}, $\Oa(t)${,} changes with time independently of the transition events, e.g., for the case of an SIR process nodes may become infected only when in contact with an infected individual {(Fig.~\ref{fig:model}A)}.
This means that the survival probability does not reduce to a simple exponential as in Eq.~(\ref{eq:S_m:static}); it is instead given by
\beq
  S_m(\tau;\ts) = \exp\left(-\int_\ts^{\tss} \Im(t)\l{m} dt \right) \es,
  \label{eq:S_m-TGA}
\eeq
where $\ts$ is the time at which the last transition took place, $\tss=\ts+\tau$ is the time when the next transition takes place, and $\Im(t)$ is an indicator function that is equal to one when the process $m$ may take place, e.g., when two given nodes are in contact, and zero when $m$ may not take place. The meaning of $\Im$ is exemplified in Fig.~{\ref{fig:model}A}: the node $i$ may {be infected by the infectious node $j$ only when the two nodes are in contact}; if we let $m$ denote this transition process, $\Im(t)$ is then one for $t=\Dt,3\Dt,4\Dt$ and zero for $t=0,2\Dt$.

{Note that for processes taking place on adaptive time-varying networks, whose changes only depend on the process itself, $\Im(t)$ only changes when a transition takes place and Eq.~(\ref{eq:S_m-TGA}) reduces to Eq.~(\ref{eq:S_m:static}). This means that from the point of view of the algorithm, such networks are effectively static and the classic ``static'' Gillespie algorithm may simply be used there~\cite{Hladish2012,Zschaler2012}.}

We now consider the general case where $\Oa(t)$ may change independently of the processes evolving on the network (as described in Sec.~\ref{sec:processes}).
Using, as in the previous section, that transition processes are independent, we can write the probability that no event takes place during an interval $\tau$ (the waiting time survival function):
\begin{eqnarray}
  S(\tau;\ts) &=& \prod_{m\in\Oa} S_m(\tau;\ts) \nonumber\\
  &=& \exp\left({-\sum_{m\in\Oa}\int_\ts^{\tss} \Im(t)\l{m} dt}\right) \es,
  \label{eq:S-TGA1}
\end{eqnarray}
where $\Oa$ {denotes} the set of all {possible} transition{s} {(transition} processes{)} on the interval {between two transition events}, $(\ts,\tss]$, i.e., $\Oa$ is the union over $\Oa(t)$ for $t\in(\ts,\tss]${, and $M$ is the total number of transition processes on the same interval (the size of $\Oa$).}
{We switch the sum and the integral in Eq.~(\ref{eq:S-TGA1}) to obtain
\beq
  S(\tau;\ts) = \exp\left({-\int_\ts^{\tss} \sum_{m\in\Oa} \Im(t)\l{m} dt}\right) \es.
\eeq
Finally, using that $\Im(t)=0$ for all $m\not\in\Oa(t)$, we may}
write
\beq
  S(\tau;\ts) = \exp\left({-\int_\ts^{\tss} \Lat \, dt}\right) \es,
  \label{eq:S(tau)}
\eeq
where
\beq
  \Lat = \sum_{{m}\in\Oa(t)} \l{m}
\eeq
is the cumulative transition rate at time $t$.

The dynamics of empirical time-varying networks is highly intermittent and we cannot describe $\Oa(t)$ analytically. This means that we cannot perform the integral of Eq.~(\ref{eq:S(tau)}) {to find the waiting time distirbution} directly.
%However, since $\Lat$ is constant except in a finite number of points of measure zero, which only depend on the network and on past transitions, and are thus known a priori,
We may instead normalize time by the instantaneous cumulative transition rate, $\Lat$:
We define {a unitless} {\sl normalized waiting time} between two consecutive transitions, $\tau'$, as
\beq
  \tau' = \L\left(\tss;\ts\right)=\int_{\ts}^{\tss} \Lat dt \es,
  \label{eq:L(t;t0)}
\eeq
i.e., equal to the cumulative transition rate integrated over $(\ts,\tss]$.
The survival function of $\tau'$ has the following simple form:
\beq
  S(\tau') = \exp({-\tau'}) \es.
\eeq
The time $\tss$ when a new transition takes place is given implicitly by $\L(\tss;\ts)=\tau'$, while the probability that $m$ is the transition that takes place at time $t=\tss$ is given by:
{\beq
  \pi_m(t) = {\Im(t)\l{m}}/{\Lat} \es.
  \label{eq:pi_m(t)}
\eeq}

This lets us define a Gillespie-type algorithm for time-varying networks by first drawing a normalized waiting time {$\tau'$} until the next event {from a standard exponential distribution [i.e. with unit rate,} $\tau'\sim\Exp(1)${]}{, and second, solving $\Lt=\tau'$ numerically to find $\tss$.}
In practice, since $\Lat$ only changes when a transition takes place or at $t_n=n\Dt$ with $n\in\mathbb{N}$, we {need only} compare $\tau'$ to
{\beq
  \L(t_{n+1};\ts) = (t_{\ns+1}-\ts)\La{\ts} + \Dt\sum_{i=\ns+1}^n \La{t_i}\es,
  \label{eq:Ln}
\eeq}
{for each time-step $n$ (Fig.~\ref{fig:algo}A--C).}
Here $\ns$ is the time-step during which the last transition took place, and $\La{t^*}$ is the cumulative transition rate at $\ts$, immediately after the last transition has taken place.
The first term of Eq.~(\ref{eq:Ln}) is the cumulative transition rate integrated over the remainder of the $\ns$th time-step left after the last transition; the second term is equal to  {$\L(t_{n+1};t_{\ns+1})$}.
A new transition takes place during the time-step $\nss$ where {${\L(t_{\nss+1};\ts)\geq\tau'}$} (Fig.~\ref{fig:algo}D); the precise time of this new transition is
{\beq
  \tss=t_\nss+\frac{\tau'-\L(t_\nss;\ts)}{\La{t_\nss}} \es{;}
  \label{eq:tss}
\eeq}
the reaction $m$ that takes place is drawn with probability given by Eq.~(\ref{eq:pi_m(t)}) {(Fig.~\ref{fig:algo}D)}.
We then update $\Oa$ and $\Lambda$ to $\Oa(\tss)$ and $\La{\tss}$ (Fig.~\ref{fig:algo}E), draw a new waiting time, $\tau'\sim\Exp(1)$, and reiterate the above procedure (Fig.~\ref{fig:algo}F).

The algorithm can be implemented for contagion processes on time-varying networks as follows (see {Methods} for pseudocode for specific contagion models and \url{https://github.com/CLVestergaard/TemporalGillespieAlgorithm} for implementation in C++):
\begin{enumerate}
  \item Draw a normalized waiting time until the first event {from a standard exponential distribution}, $\tau'\sim\Exp(1)$ {(Fig.~\ref{fig:algo}A)}.
  \item At each time-step $t_n=n\Dt$, with $n=0,1,2,\ldots$, let $\Oa\equiv\Oa(t_n)$ and $\Lambda\equiv\La{t_n}$; here, only contact-dependent processes (type {\sl b}, Sec.~\ref{sec:compartmental}) and mixed (type {\sl c}, Sec.~\ref{sec:compartmental}) processes that depend on contacts taking place at $t_n$ or $t_{n-1}$ need to be updated---an important point, as it lets the temporal Gillespie algorithm be {much} faster than rejection sampling (see discussion in Sec.~\ref{sec:comparison}).
      Then, compare $\tau'$ to $\Lalgo$:
    \begin{description}
      \item[if $\Lalgo\leq\tau'$] Subtract $\Lalgo$ from $\tau'$, continue to next time-step and repeat 2 {(Figs.~\ref{fig:algo}A--\ref{fig:algo}C)} {\cite{footnote2}}.
      \item[if $\Lalgo>\tau'$] Let the reaction $m$ take place, chosen from $\Oa$ with probability $\pi_m=\l{m}/\Lambda$. The fraction that is left of the time-step when the transition takes place is $\xi=1-\tau'/(\Lalgo)$ and the precise time of the transition is $\tss=t_n+\tau'/\Lambda$ {(Figs.~\ref{fig:algo}D and \ref{fig:algo}F)}.
          Next, update $\Oa$ and $\Lambda$ {(Fig.~\ref{fig:algo}E)}; this time all transition processes should be updated, as spontaneous processes (type {\sl a}, Sec.~\ref{sec:compartmental}) may change, emerge, or disappear when a transition takes place.
          Then:
          \begin{enumerate}[label=(\alph*)]
            \item draw a new normalized waiting time, $\tau'\sim\Exp(1)$ {(Fig.~\ref{fig:algo}F)};
            \item compare $\tau'$ to $\xi\Lalgo$:
              \begin{description}
                \item[if $\tau'\geq\xi\Lalgo$] subtract $\xi\Lalgo$ from $\tau'$, continue to the next time-step and repeat 2 {(Fig.~\ref{fig:algo}F)}.
                \item[if $\tau'<\xi\Lalgo$] Another transition takes place during the present time-step (at time $t^{***}=\tss+\tau'/\Lambda$, where $\tss$ is the time of the last transition during the same time-step): choose $m$ from $\Oa$ with probability $\pi_m=\l{m}/\Lambda$; let $\xi\to\xi-\tau'/\Lalgo$, and update $\Oa$ and $\Lambda$.
                    Repeat a) and b).
              \end{description}
          \end{enumerate}
    \end{description}
\end{enumerate}

By construction, the above procedure produces realizations of a stochastic process for which the waiting times for each transition follow exactly their correct distributions.
The temporal Gillespie algorithm is thus what we term {\sl stochastically exact}: all distributions and moments of a stochastic process evolving on a time-varying network obtained through Monte Carlo simulations converge to their exact values. Rejection based sampling algorithms are stochastically exact only in the limit $\l{m}\Dt\to0$.

A large literature exists on the related problem of simulating coupled chemical reactions under externally changing conditions (e.g., time-varying temperature or volume)~\cite{Gibson2000,Lu2004,Anderson2007,Carletti2012,Caravagna2013,Zechner2014}. Most of these methods consider only external perturbations that can be described by an analytical expression. In this case the problem reduces to that of defining a static, yet non-Markovian, algorithm.
Some methods, and notably the {\sl modified next reaction method} developed by Anderson~\cite{Anderson2007}, can be adapted to a completely general form of the external driving and thus, in principle, to simulate dynamical processes taking place on time-varying networks.
These methods are based on a {scheme} that is conceptually similar to Gillespie's direct algorithm, the {\sl next reaction method}, proposed by Gibson and Bruck~\cite{Gibson2000}.
The next reaction method draws a waiting time for each reaction individually and chooses the next reaction that happens as the one with the shortest corresponding waiting time. It then updates the remaining waiting times, draws new waiting times (if applicable), and reiterates.
To generalize the next reaction method to processes with non-exponential waiting times, Anderson introduced the concept of the {\sl internal time} for each transition process~\cite{Anderson2007}.
In the notation used in the present article it is defined as $T_m(t)=\int_0^t\Im(t)\lambda_m dt$ and is thus equivalent to the normalized time, $\L(t,0)$, only for an individual transition process.

By construction, the next reaction method needs to draw only one random number per transition event, where the Gillespie algorithm draws two. However, this reduction in the number of required random variables comes at a price: one must {draw a random number for each individual transition process and} keep track of, compare, and update each of the individual waiting times. For chemical reactions, where the number of different chemical reactions is small (it scales with the number of chemical species), this tradeoff favors the next reaction method.
However, for contagion processes on networks, each individual is unique (if not intrinsically, at least due to its position in the network).
The number transition processes thus scales with the number of nodes and contacts, which favors the Gillespie algorithm as it does not need to keep track of each of them individually~\cite{Boguna2014}.

On time-varying networks (or for time-varying external driving) one must furthermore update relevant internal times each time the network structure (external conditions) changes in the next reaction method.
Chemically reacting systems are usually close to being adiabatic, i.e., the external driving changes slowly compared to the time-scales of chemical reactions. Thus, the additional overhead related to updating individual internal times is practically negligible.
However, the dynamics of temporal networks is highly intermittent and the time-scale of network change is typically smaller than the time-scales of relevant dynamical processes. Here one must thus update the internal times many times between each transition event, inducing a substantial overhead.
Since the temporal Gillespie algorithm operates with a single global normalized waiting time, it handles these updates more efficiently.

Finally, the modified next reaction method may in principle be extended to non-Markovian processes taking place on time-varying networks (as treated in Sec.~\ref{sec:nonMarkovian} {using the temporal Gillespie algorithm}).
However, such an approach would, for each single transition, require solving numerically Eq.~(13) of \cite{Anderson2007} for the internal waiting time of each individual transition process, taking into account the time-varying network structure, finding the shortest corresponding waiting time in real time, and then updating the internal waiting times of all the other reactions, rendering the next reaction method even more inefficient in this general case.

\subsection{Comparison of Gillespie and rejection sampling algorithms}
\label{sec:comparison}
\begin{figure*}
%\begin{adjustwidth}{-2.25in}{0in} % Comment out/remove adjustwidth environment if table fits in text column.
   \includegraphics{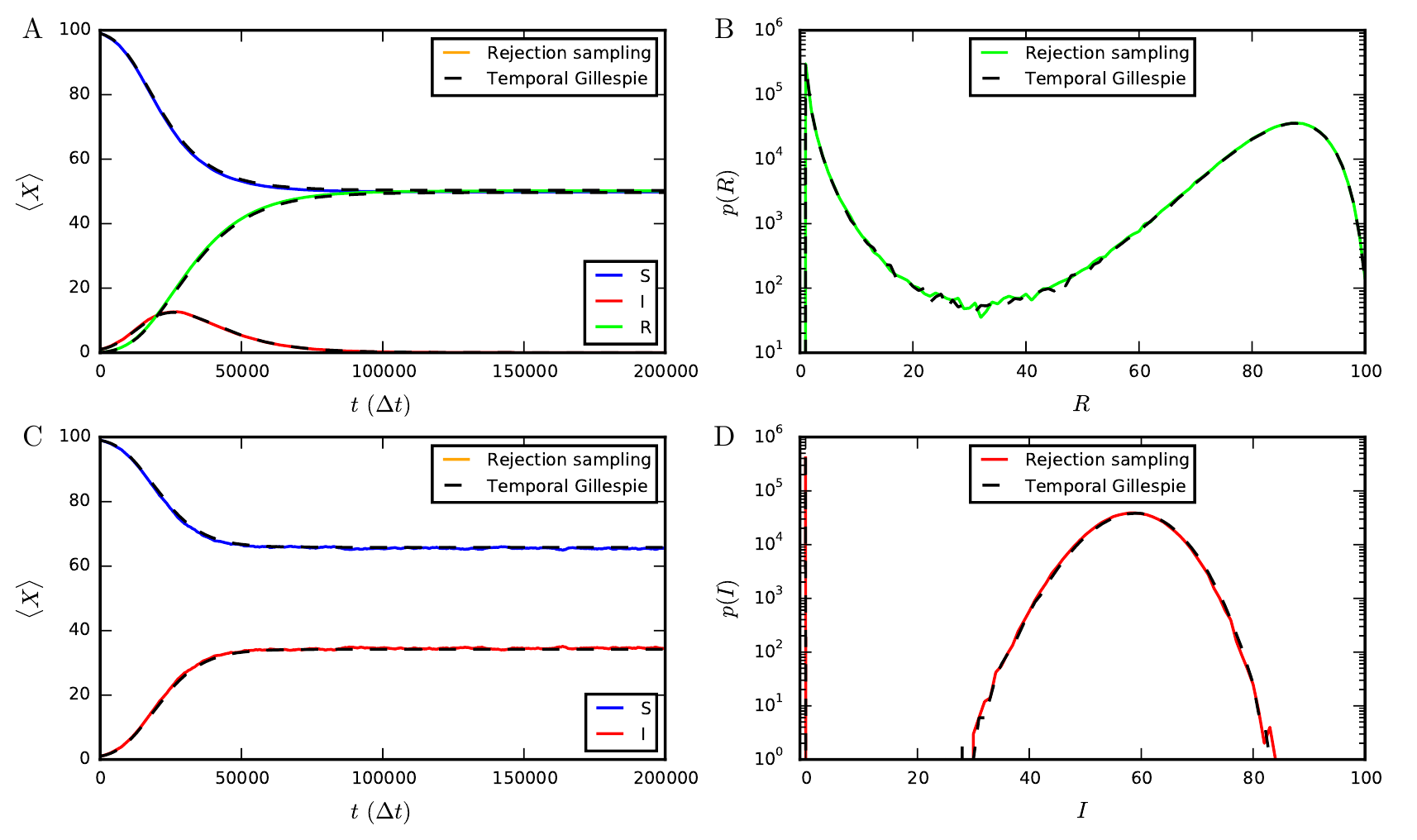}
   \caption{{\bf Comparison of numerical results from temporal Gillespie and rejection sampling algorithms.}
   (A) Mean number of nodes in each state of the SIR model as function of time.
   (B) Distribution of final epidemic sizes (number of recovered nodes when $I=0$) in the SIR model.
   (C) Mean number of nodes in each state of the SIS model as function of time.
   (D) Distribution of the number of infected nodes in the stationary state ($t\to\infty$) of the SIS model.
   All simulations were performed 1\,000\,000 times with the root node chosen at random on an activity driven network~\cite{Perra2012} consisting of $N=100$ nodes, with activities $a_i=\eta z_i$, where $\eta=0.1$ and $z_i\sim z_i^{-3.2}$ for $z_i\in[0.03,1)$, and a node formed two contacts when active.
   Parameters of the epidemic processes were $\beta\Dt=10^{-2}$ and $\mu\Dt=10^{-4}$.}
   \label{fig:validate}
%\end{adjustwidth}
\end{figure*}

\paragraph*{Numerical validation.}
We compare the outcome of SIR and SIS processes on activity-driven time-varying networks~\cite{Perra2012} simulated using the temporal Gillespie algorithm to {the outcome of} simulations using traditional rejection sampling.
For sufficiently small $\l{m}\Dt$, the outcomes are indistinguishable (Fig.~\ref{fig:validate}{, see also Supplementary Fig.~\ref{S1_Fig} for an empirical network of face-to-face contacts in a high school}), confirming the validity of the temporal Gillespie algorithm. Note that rejection sampling is only expected to be accurate for $\l{m}\Dt\ll1$, while the temporal Gillespie algorithm is stochastically exact for all $\l{m}\Dt$; the results of the two algorithms thus differ when the assumption $\l{m}\Dt\ll1$ does not hold (Supplementary Fig.~\ref{S2_Fig}).

\begin{figure*}
%\begin{adjustwidth}{-2.25in}{0in} % Comment out/remove adjustwidth environment if table fits in text column.
   \centering
   \includegraphics{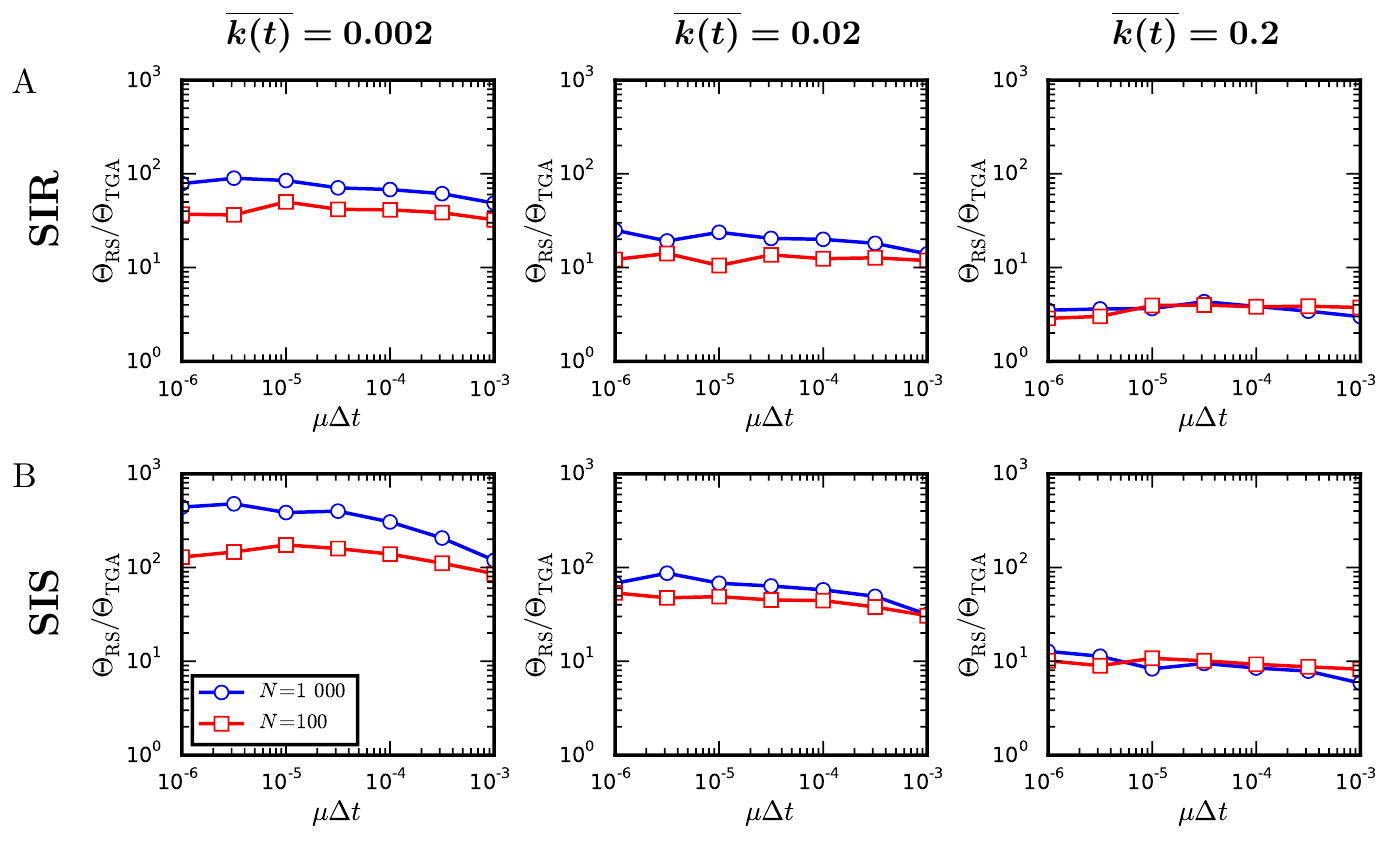}
   \caption{{\bf Comparison of the speed of the temporal Gillespie and the rejection sampling algorithms.}
   {Ratio between computational times $\Trs$ and $\Ttga$ per single realization of a spreading process using rejection sampling and the temporal Gillespie algorithm, respectively:
   (A) for a SIR process and (B) for a SIS process on networks with different mean degree, $\kt$} [for empirical contact networks, $\kt\approx0.004$--$0.07$ ({Table~\ref{tab:empiricalNetworks}})].
   Networks consisted of $N=100$ or $N=1000$ nodes, with activities $a_i=\eta z_i$ and $z_i\sim z_i^{-3.2}$ for $z_i\in[0.03,1)$; a node formed two contacts each time it was active.
   For $\Dt=20\,{\rm s}$ (as for the empirical data), $\mu\Dt\approx3\cdot10^{-5}$ corresponds to a recovery time of roughly one week, typical of flu-like diseases.
   The infection rate was $\beta=10^{3}\mu$ for networks with $\kt=0.002$, $\beta=10^{2}\mu$ for networks with $\kt=0.02$, and $\beta=10\mu$ for networks with $\kt=0.2$.
   {(Details on how simulations were performed are found in {Methods}.)}}
   \label{fig:speed}
%\end{adjustwidth}
\end{figure*}

\begin{figure*}
%\begin{adjustwidth}{-2.25in}{0in} % Comment out/remove adjustwidth environment if table fits in text column.
   \centering
   \includegraphics{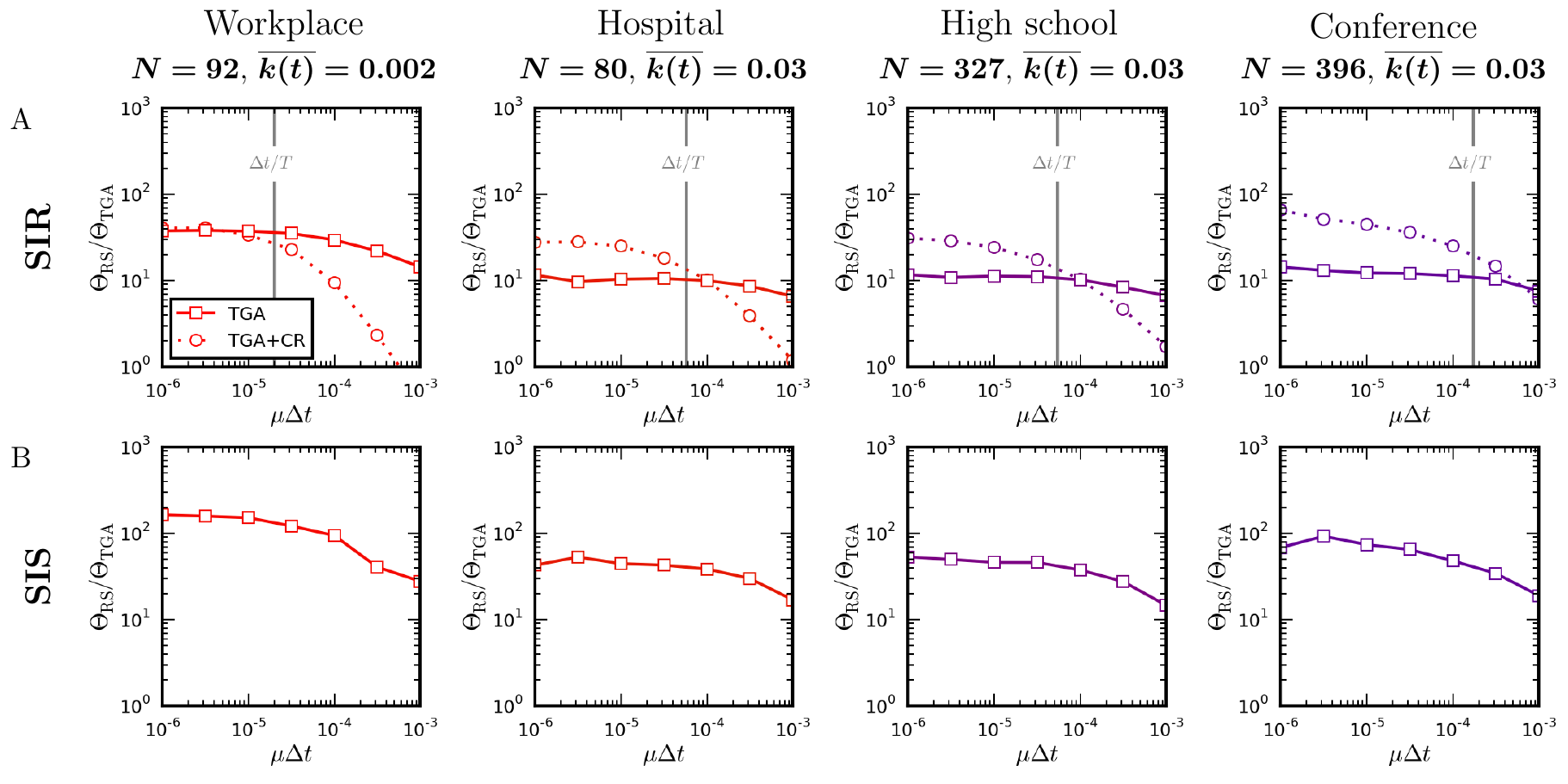}
   \caption{{{\bf Comparison of the speed of the temporal Gillespie and rejection sampling algorithms on empirical time-varying networks.}
   Ratio $\Trs/\Ttga$ between the time per realization of a single simulation using rejection sampling and the temporal Gillespie algorithm on empirical face-to-face contact networks {in different social settings (Table~\ref{tab:empiricalNetworks})}:
   {(A) for a SIR process, without (TGA) and with (TGA+CR) contact removal; (B) for a SIS process.}
   Simulations were performed with $\beta=1\,000\mu$ for the workplace and $\beta=100\mu$ for the other networks.
   (Details on how simulations were performed are found in {Methods}.)}}
   \label{fig:empirical}
%\end{adjustwidth}
\end{figure*}

\paragraph*{Comparison of simulation speed.}
Next, we compare the speeds of the temporal Gillespie and the rejection sampling algorithms for SIR and SIS processes (see {Methods} for details on how simulations were performed). Figure~\ref{fig:speed} shows that the temporal Gillespie algorithm is up to multiple orders of magnitude faster than traditional rejection sampling. These results are confirmed by simulations on empirical time-varying networks {of face-to-face contacts}~({Fig.~\ref{fig:empirical}, Table~\ref{tab:empiricalNetworks}}).
The speed gain is higher for larger systems (compare $N=1\,000$ to $N=100$ in Fig.~\ref{fig:speed})
We also see that the speed gain is larger the sparser the network is. This is because the calculation of the contacts between susceptible and infected nodes at each time-step, necessary to determine the possible $\S\to\I$ transitions, is the performance limiting step of the temporal Gillespie algorithm {(see below)}. In the extreme case of a contagion model where all transitions are contact-dependent (type {b}, Sec.~\ref{sec:processes}), such as the classic Maki-Thompson model of rumor spreading~{\cite{Maki1973}}, the temporal Gillespie algorithm is approximately a factor two faster than the rejection sampling algorithm.
\begin{table}
  \caption{{\bf Summary statistics for empirical face-to-face contact networks from the {\em SocioPatterns} collaboration~\cite{sociopatterns}:} social setting; number of nodes in the network, $N$; total duration of measurements, $T$; average instantaneous degree, $\kt$.}
  \label{tab:empiricalNetworks}
  \begin{tabular}{|l|c|c|c|c|}
    \hline
    {\bf Setting} & $\bm N$ & $\bm T$ & $\bm{\kt}$ & {\bf Reference}\\
    \hline
    Workplace & 92 & 11 days & 0.004 & \cite{Genois2015} \\
    \hline
    Hospital & 80 & 4 days & 0.064 & \cite{Vanhems2013} \\
    \hline
    High school & 327 & 4 days & 0.063 & \cite{Fournet2014} \\
    \hline
    Conference & 399 & 32 hours & 0.070 & \cite{Stehle2011} \\
    \hline
  \end{tabular}
\end{table}

\paragraph*{Expected time complexity of the algorithms.}
{We may gain insigth into the performance of the algorithms by considering their {\sl time-complexity}, i.e., how their running time scales with the input parameters of the simulated system. Since the algorithms are used for Monte Carlo simulations, it is most interesting to consider the {\sl expected complexity} given a set of parameters, i.e., the mean running time of an algorithm averaged over an ensemble of simulations, not the {\sl worst-case complexity} which is usually considered for deterministic algorithms.}

{The expected running time of the rejection sampling algorithm scales as
\beq
  \Trs = \Ord{\Emean\,\nsimu} + \Ord{\Mmean\,\nsimu} \es,
  \label{eq:T_RS}
\eeq
where $\Ord{x}$ denotes a term that is of order $x$, $\Emean=N\kt/2$ is the mean number of contacts per time-step, $\Mmean$ is the mean number of possible transitions at any instant, and $\nsimu$ is the number of time-steps simulated.
For comparison, the expected running time of the temporal Gillespie algorithm is given by
\beq
  \Ttga = \Ord{\Emean\,\nsimu} + \Ord{\Qmean\,\nsimu} \es,
  \label{eq:T_TGA}
\eeq
where $\Qmean$ is the mean number of transitions that take place per time-step.}

{The first term of the r.h.s. of Eqs.~(\ref{eq:T_RS}) and (\ref{eq:T_TGA}) correspond to the time needed for looking through the set of contacts at each time-step to determine the set of possible infections and are thus similar for the rejection sampling and temporal Gillespie algorithms (with the temporal Gillespie algorithm incurring a small additional overhead related to calculating the cumulative transition rate and keeping track of of the normalized waiting time left till the next transition). For rejection sampling [Eq.~(\ref{eq:T_RS})], the second term corresponds to the determination of whether each of the possible transitions takes place at each time-step; for the temporal Gillespie algorithm [Eq.~(\ref{eq:T_TGA})], the second term corresponds to drawing inter-event waiting times and which transitions that take place.
For the SIR and SIS processes considered above, $\Mmean=\Msi+\Imean$, where $\Msi$ is mean the number of possible $S\to I$ transitions per time-step, and $\Imean$ is the mean number of infected nodes.}

{Empirically relevant networks are sparse and transition rates are small, so typically $\Qmean\ll\Emean\ll\Mmean$. (The first inequality is a consequence of transition rates being small compared to $1/\Dt$; the second inequality follows by noting that $\Imean\sim N\gg\Emean$.)
This means that the performance of the rejection sampling algorithm is limited by the rejection sampling step [second term of Eq.~(\ref{eq:T_RS})], while the performance of the temporal Gillespie algorithm is limited by the iteration over the set of contacts in order to update $\Oa(t)$ [first term in Eq.~(\ref{eq:T_TGA})]; this explains why the difference in performance decreases with the mean instantaneous degree of the network.
This also hints at how we may improve the speed of the temporal Gillespie algorithm: by rendering the identification of relevant contacts during each time-step faster.
One such approach which may be applied to processes with an absorbing state (e.g. an $\R$ state) is explored below. }

\paragraph*{{Improving performance by removing obsolete contacts.}}
{Empirical networks describing human contact differ from simulated networks in a number of ways. For example, their structure and dynamics are more complex~\cite{Gauvin2013,Stehle2011,Vanhems2013,Fournet2014,Genois2015} but perhaps most importantly in the perspective of optimizing simulations, they are of finite length.
One is often interested in long-time behavior or slowly evolving processes compared to the length of available data. To overcome this limitation, one usually loops over the data set. This means that if a node enters an inactive absorbing state such as the recovered ($\R$) state in the SIR model, one may remove all following contacts to this node from the data, thus reducing the number of contacts that one must go through during the following loop. Furthermore, since the $\I\to\R$ transition is independent of the network, one may also remove all contacts between two infected nodes.}

{Pseudocode for an algorithm that removes obsolete contacts is given in {Methods} and C++ code can be found at~\url{https://github.com/CLVestergaard/TemporalGillespieAlgorithm}. Figure~\ref{fig:empirical}(A) compares the speed gain of the temporal Gillespie algorithm relative to rejection sampling with and without contact removal for simulations of a constant-rate SIR process on empirical networks of face-to-face contacts (Table~\ref{tab:empiricalNetworks}).
Depending on the parameters of the simulated process, removing obsolete contacts may induce both a significant gain or loss in speed; for processes that are fast compared to the length of the data set, the data is not repeated or only repeated few times during a simulation and the additional overhead involved in identifying and removing the obsolete contacts renders the algorithm slower; for slow processes the data is looped many times and removing the obsolete contacts makes the algorithm faster.
Figure~\ref{fig:empirical}(A) suggests an empirically determined rule-of-thumb: if the slowest time-scale of the simulated process (here $\sim1/\mu$) is longer than the length of the data, $T$, removing obsolete contacts pays off, if it is shorter, one should not remove obsolete contacts.}

\paragraph*{{Slow network dynamics.}}
{For time-varying networks of face-to-face contacts, which are relevant for simulating epidemic spreading in a population, network dynamics are typically much faster than the time-scales of the dynamical process that is simulated (compare the $20\,{\rm s}$ time-resolution of the empirical data of Table~\ref{tab:empiricalNetworks} to typical $1/\beta\sim1\,$hour and $1/\mu\sim1\,$week for flu-like diseases).
In the opposite case, i.e., if the network evolves much slower than the dynamical process, the temporal Gillespie algorithm simply works like a static Gillespie algorithm in-between changes in the network structure while taking the changes changes into account exactly when they occur. The performance of the temporal Gillespie algorithm then approaches that of a static Gillespie algorithm in this case. Note that since $\Qmean\gg\Emean$ in this limit, the second term dominates in Eq.~(\ref{eq:T_TGA}), which means that the speed of the algorithm is limited by the selection of waiting times and transitions that take place, and care should be taken to optimize these steps, e.g., by organizing the transition processes in a heap or a priority queue~\cite{Anderson2007}.} %Drawing a waiting time is an $\Ord{1}$ operation, while the complexity of choosing the transition that takes place depends on both implementation and the }
{Note finally that to obtain reliable results using a rejection sampling algorithm one must use a time-step for simulations $\Dt_{\rm RS}$ which is much smaller than the time-step $\Dt$ of network change. Thus the expected time complexity of rejection sampling scales with $\Dt/\Dt_{\rm RS}\,\nsimu\gg\nsimu$ in this case.}

\subsection{Non-Markovian processes}
\label{sec:nonMarkovian}
For real-world contagion processes, transition rates are typically not constant but in general depend on the history of the process~\cite{Ferguson2006,Lloyd2001}. Such processes are termed {\sl non-Markovian}.
The survival probability for a single non-Markovian transition process taking place on a time-varying network is given by:
\beq
  S_m\left(\tau;\F{m}\right) = \exp\left({-\int_{\ts}^{\tss}{\rm I}_m(t)\l{m}\left(t;\F{m}\right)dt}\right) \es.
  \label{eq:S_m-nM}
\eeq
%\begin{eqnarray}
%  S_m\left(\tau_m;\F{m}\right)&=&\prod_\t{m}^{\t{m}+\tau_m}\left[1-{\rm I}_m(t)\l{m}\left(t;\F{m}\right)dt\right] \\
%  &=& \exp\left[{-\int_{\t{m}}^{\t{m}+\tau_m}{\rm I}_m(t)\l{m}\left(t;\F{m}\right)dt}\right]  \nonumber\es.
%\end{eqnarray}
Here $\F{m}$ is a filtration for the process $m$, i.e., all information relevant to the transition process available up to and including time $t$; typically, $\F{m}$ will be its starting time and relevant contacts that have taken place since.
As in Sec.~\ref{sec:temporalGillespie}, $\ts$ is the time of the last transition and $\tss=\ts+\tau$ is the time of the next.
[Note that since $\l{m}$ now depends explicitly on $t$, we may absorb $\Im$ in $\l{m}$; however, to underscore the analogy with the Poissonian case, we keep the factor $\Im$ explicitly in Eq.~(\ref{eq:S_m-nM}).]

We use again that the transition processes are independent, to write the waiting time survival probability:
\begin{eqnarray}
  S\left(\tau;\Ft\right) %&=& \prod_{m\in\Oa} S_m\left(\tau;\F{m}\right) \nonumber\\
%  &=& \exp\left({-\sum_{m\in\Oa}\int_0^{\tau}{\rm I}_m(t)\l{m}\left(t;\F{m}\right)dt}\right) \nonumber\\
  &=& \exp\left({-\int_\ts^{\tss}\Lambda\left(t;\Ft\right)dt}\right) \label{eq:S_nonMarkov} \es,
\end{eqnarray}
with
\beq
  \Lambda\left(t;\Ft\right) = \sum_{m\in\Oa(t)}\l{m}\left(t;\F{m}\right) \es,
\eeq
and where $\Ft$ is the union over $\F{m}$ for $m\in\Oa$.

For a static network{,} Eq.~(\ref{eq:S_nonMarkov}) reduces to the result found in~\cite[Eq.~(7)]{Boguna2014}. This can be seen by noting that $M(t)=M$ and $\Oa(t)=\Oa$ are then constant, and thus that $\l{m}(t;\F{m})=-[dS_m(t;\F{m})/dt]/S_m(t;\F{m})=d\{\ln[1/S_m(t;\F{m})]\}/dt$ and $S_m(t;\F{m})=S_m(t+t_m;\F{m})/S_m(t_m;\F{m})$, yielding directly Eq.~(7) of~\cite{Boguna2014}.

As in the Poissonian case (Sec.~\ref{sec:temporalGillespie}) we define the normalized waiting time, $\tau'$, as
\beq
  \tau' = \L(\tss;\ts,\Ft)=\int_{\ts}^\tss \La{t;\Ft}dt \es.
\eeq
This gives us the same simple forms as above for the survival function of the normalized waiting time, $\tau'$,
\beq
  S(\tau') = \exp(-\tau') \es,
\eeq
and the probability that $m$ is the transition that takes place at $t=\tss$,
\beq
  {\pi_m(t;\Ft) = \Im(t) \frac{\l{m}\left(t;\F{m}\right)}{\Lambda\left(t;\Ft\right)} \es.}
\eeq

Until now our approach and results are entirely equivalent to the Poissonian case considered above.  However, since $\l{m}(t)$ in general depend continuously on time, the transition time $\tss$ is not simply found by linear interpolation as in Eq.~(\ref{eq:tss}). Instead, one would need to solve the implicit equation $\L(\tss;\ts)=\tau'$ numerically to find $\tss$ exactly.
To keep things simple and speed up calculations, we may approximate $\Lat$ as constant over a time-step. This assumes that $\Delta\Lat\Dt\ll1$, where $\Delta\Lat$ is the change of $\Lat$ during a single time-step. It is a more lenient assumption than the assumption that $\Lat\Dt\ll1$ which rejection sampling relies on, as can be seen by noting that in general $\Delta\Lat/\Lat\ll1$.
The same assumption also lets us calculate {$\L(t_{n+1};\ts)$} as in the Poissonian case:
\beq
  {\L(t_{n+1};\ts,\Ft) = (t_{\ns+1}-\ts)\Lambda(\ts) + \Dt\sum_{i=\ns+1}^n\La{t_i;\Ft} \es,}
\eeq
and the time, $\tss$, at which the next transition takes place:
\beq
  {\tss=t_\nss+\frac{\tau'-\L(t_\nss;\ts,\Ft)}{\La{t_\nss;\Ft}} \es.}
\eeq
Using the above equations, we can now construct a temporal Gillespie algorithm for non-Markovian processes.

\begin{figure*}
%\begin{adjustwidth}{-2.25in}{0in}
   \includegraphics{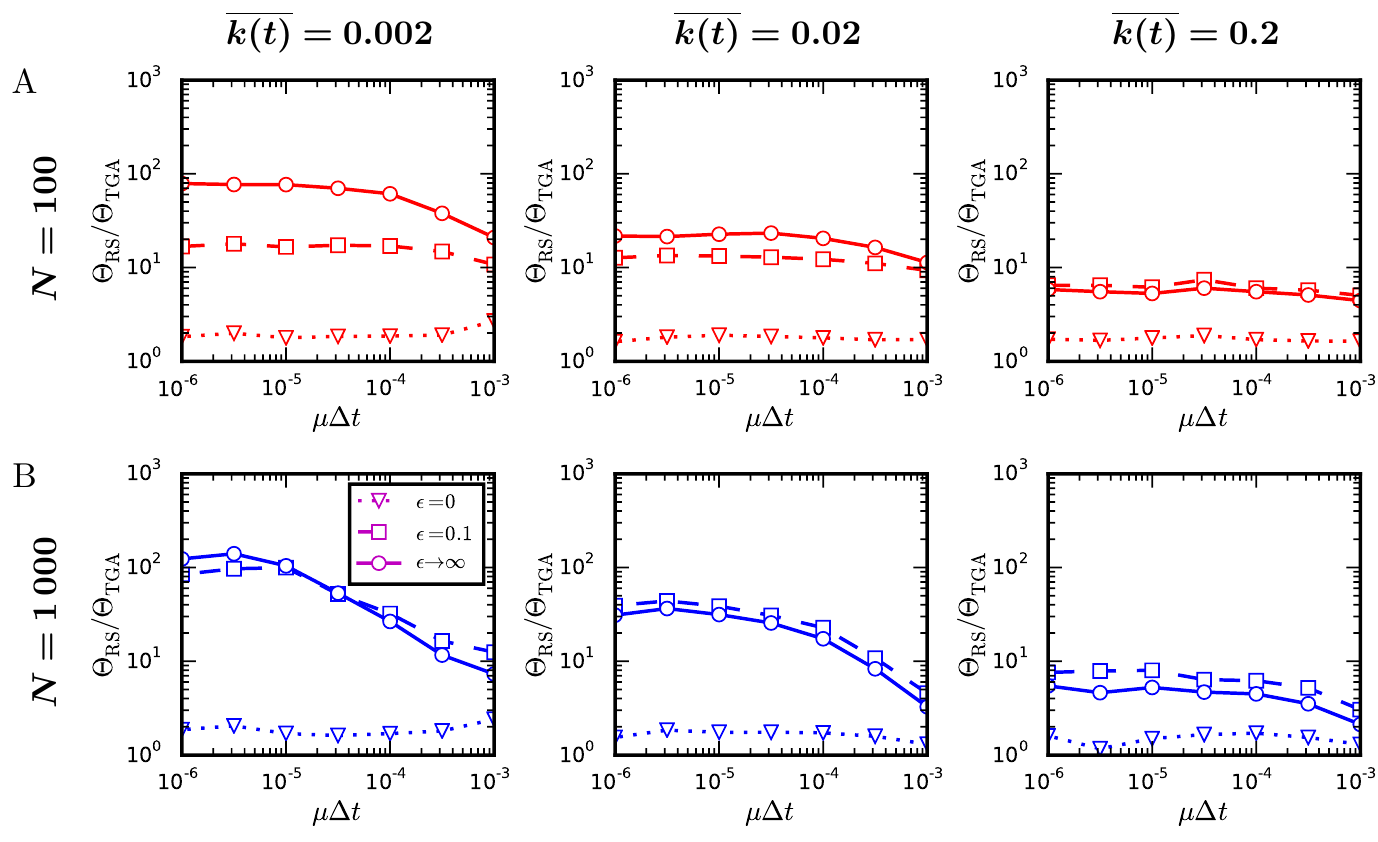}
   \caption{{\bf Comparison of the speed of the temporal Gillespie and the rejection sampling algorithms: non-Markovian SIR process.}
   {Ratio $\Trs/\Ttga$ between the} time per realization of a single simulation of an SIR process with Weibull distributed recovery times {using rejection sampling and the temporal Gillespie algorithm} on activity driven network{s of different average degree $\kt$ [for empirical contact networks, $\kt\approx0.004$--$0.07$ (Table~\ref{tab:empiricalNetworks})]}:
   {(A) for networks consisting of $N=100$ nodes and (B) of $N=1\,000$ nodes.}
   The parameter $\epsilon$ controls the accuracy of the temporal Gillespie algorithm: for $\epsilon=0$, where $\l{m}(t)$ is approximated as constant over a single time-step, it is most accurate; for $\epsilon\to\infty$, where $\l{m}$ is approximated as constant between two consecutive transition events, it is the least accurate.
   {Node} activities {were given by} $a_i=\eta z_i$ {with} $z_i\sim z_i^{-3.2}$ for $z_i\in[0.03,1)$; a node formed two contacts each time it was active.
   The recovery rate of an infected node was given by Eq.~(\ref{eq:Weibull}) with $\gamma=1.5$.
   The infection rate was $\beta=10^{3}\mu_0$ for networks with $\kt=0.002$, $\beta=10^{2}\mu_0$ for networks with $\kt=0.02$, and $\beta=10\mu_0$ for networks with $\kt=0.2$.
   {(See {Methods} for details on how simulations were performed.)}}
   \label{fig:speedSIR-nM}
%\end{adjustwidth}
\end{figure*}

\begin{figure*}
%\begin{adjustwidth}{-2.25in}{0in}
   \includegraphics{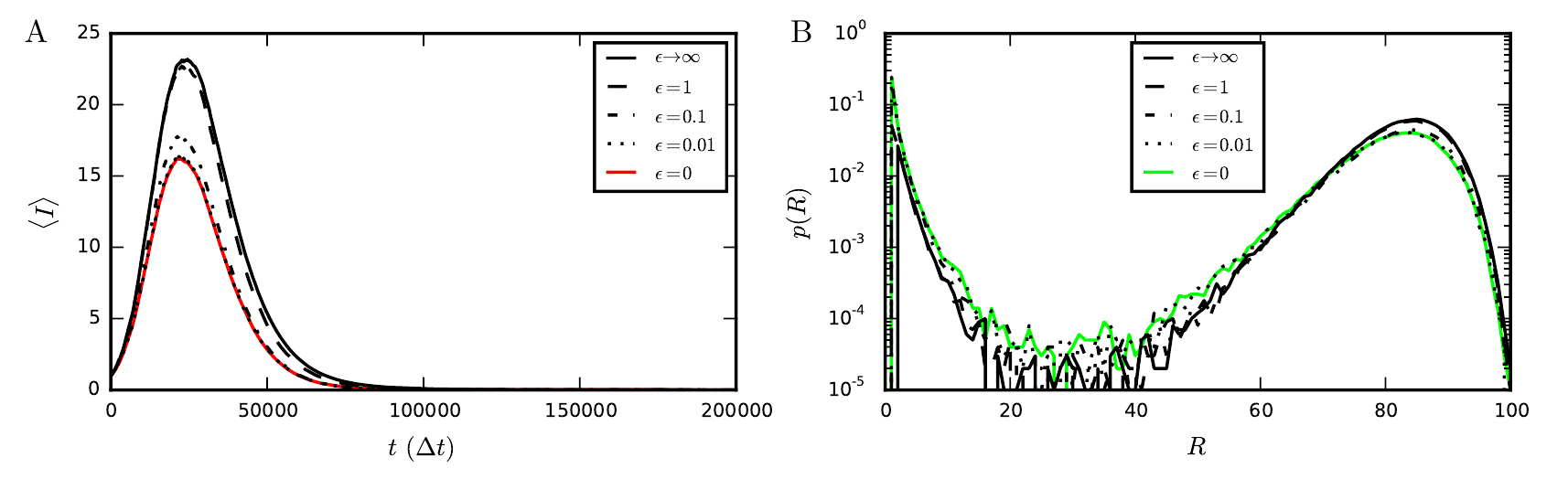}
   \caption{{\bf Comparison of the outcome of non-Markovian SIR processes for different values of the parameter $\epsilon$.}
   (A) Average number of infected nodes, $\langle I\rangle$, as function of time for a SIR process with Weibull distributed recovery times.
   (B) Distribution of the numbers of recovered nodes after the infection has died out (i.e. when $I=0$).
   For $\epsilon=0$ the temporal Gillespie algorithm is quasi-exact (see Supplementary Fig.~\ref{S3_Fig} for comparison with rejection sampling); for $\epsilon\to\infty$, corresponding to a first-order cumulant expansion of $\La{t;\Ft}$ around $t=\ts$ (see main text), it is least accurate.
   As $\epsilon$ is decreased, both $\langle I\rangle(t)$ and $p(R)$ rapidly approach the quasi-exact result obtained for $\epsilon=0$.
   Simulations were performed on an activity-driven network consisting of $N=100$ nodes with activities $a_i= z_i/10$, where $z_i\sim z_i^{-3.2}$; nodes' recovery times followed Eq.~(\ref{eq:Weibull}) with $\gamma=1.5$ and $\mu_0=10^{-4}\,{\rm Hz}$; the length of a time-step was $\Dt=1\,{\rm s}$ and the infection rate $\beta=100\mu_0=10^{-2}\,{\rm Hz}$. }
   \label{fig:validate-nM}
%\end{adjustwidth}
\end{figure*}

This algorithm updates all $\l{m}(t)$ that depend on time at each time-step, where for the Poissonian case we only had to initialize new processes, i.e., contact-dependent processes (type {\sl b} and {\sl c}, Sec.~\ref{sec:compartmental}). This means the algorithm is only roughly a factor two faster than rejection sampling [compare dotted lines ($\epsilon=0$) in Fig.~\ref{fig:speedSIR-nM}].
To speed up the algorithm, we may employ a first-order cumulant expansion of $S(\tau;\Ft)$ around $\tau=0$, as proposed in~\cite{Carletti2012,Boguna2014} for static non-Markovian Gillespie algorithms. It consists in approximating $\l{m}(t;\F{m})$ by the constant $\l{m}(\ts;\F{m})$ for $\ts<t<\tss$ and gives a considerable speed increase of the algorithm [full lines ($\epsilon\to\infty$) in Fig.~\ref{fig:speedSIR-nM}]. However, the approximation is only valid when $M(t)\gg1$~\cite{FirstOrderCumulant}, which is not always the case for contagion processes. Notably, at the beginning and end of an SIR process, and near the epidemic threshold for an SIS process, $M$ is small and the approximation breaks down; the approximate algorithm for example overestimates the peak number of infected nodes in a SIR process with recovery rates that increase over time [compare full black line ($\epsilon\to\infty$) to the quasi-exact full red line ($\epsilon=0$) in Fig.~\ref{fig:validate-nM}A]. %In particular
An intermediate approach, which works when the number of transition processes is small, but is not too slow to be of practical relevance, is needed. We propose one such approach below~\cite{SecondOrderCumulant}.

\paragraph*{Efficient non-Markovian temporal Gillespie algorithm.}
\label{sec:nonMarkov-algo}
As discussed above, we neither want to update all transition rates at each time-step as this makes the temporal Gillespie algorithm slow, nor do we want to only update them when a transition event takes place as this makes the algorithm inaccurate.

An intermediate approach is found by looking at the relevant physical time-scales of the transition processes: the average waiting time before they take place, $\taum$.
If the time elapsed since we last updated $\l{m}(t)$ is small compared to $\taum$, we do not make a large error by treating it as constant over the interval; however, if the elapsed time is comparable to or larger than $\taum$, the error may be considerable.
Thus, instead of updating $\l{m}$ at each time-step, we may update it only after a time $t>\epsilon\taum$ has elapsed since it was last updated.
Here $\epsilon$ controls the precision of the algorithm.

{Below,} we use this approach to simulate a non-Markovian SIR process, where the recovery times of infected nodes follow a Weibull distribution (see {Methods} for an algorithm written in pseudocode and \url{https://github.com/CLVestergaard/TemporalGillespieAlgorithm} for implementation in C++).
The recovery rate of an infected node is here given by
\beq
  \mu(t;\t{m}) = \gamma\mu_0^\gamma\left(t-\t{m}\right)^{\gamma-1} \es,
  \label{eq:Weibull}
\eeq
where $\mu_0$ sets the scale, $\t{m}$ is time when the node was infected, and $\gamma$ is a shape parameter of the distribution. For $\gamma=1$, we recover the constant-rate Poissonian case with $\mu(t;\t{m}) = \mu_0$. For realistic modeling of infections, $\gamma>1$; here $\mu(t;\t{m})$ is zero at $t=\t{m}$ and grows with time.
In this case, we thus update the recovery rates $\mu(t;\t{m})$ whenever the time elapsed since a transition last took place exceeds $\taum=\Gamma(1+1/\gamma)/\mu_0$.

\begin{figure}
%\begin{adjustwidth}{-2.25in}{0in}
   \includegraphics{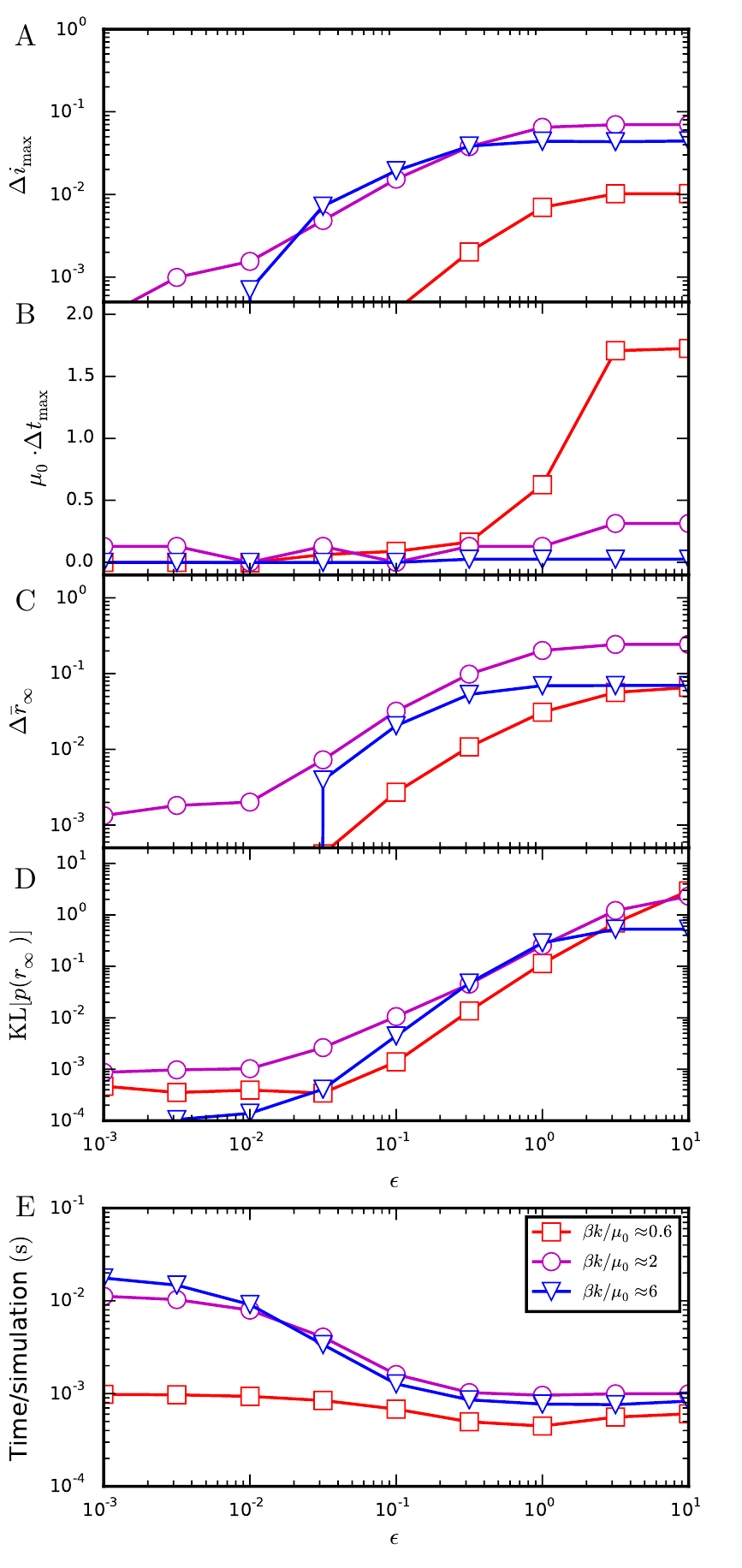}
   \caption{{\bf Accuracy and speed of the non-Markovian temporal Gillespie algorithm as function of $\epsilon$.}
   (A)--(D) Different measures of the difference in outcome of simulations between algorithms with $\epsilon>0$ and $\epsilon=0$ (quasi-exact).
   (A) Difference, $\Delta \overline{i}_{\rm max}=\overline{i}_{\rm max}(\epsilon)-\overline{i}_{\rm max}(0)$, in the peak average fraction of infected nodes,  $\overline{i}_{\rm max}=\overline{I}_{\rm max}/N$.
   (B) Difference $\Delta t_{\rm max}$ between the times at which this peak takes place, normalized by $\mu_0$.
   (C) Difference, $\Delta \overline{r}_{\infty}$, in the average fraction of nodes affected by the infection---the average {\sl attack rate}.
   (D) Kullback-Leibler divergence, ${\rm KL}[p(r_{\infty})]$, between the distributions of attack rates
   (E) Time per simulation of the process.
   Simulations were performed on an activity-driven network with $N=100$ nodes and activities $a_i= z_i/10$ with $z_i\sim z_i^{-3.2}$ for $z_i\in[0.03,1)$; nodes' recovery times followed Eq.~(\ref{eq:Weibull}) with $\gamma=1.5$, $\mu_0=10^{-4}\,{\rm Hz}$, and $\Dt=1\,{\rm s}$.}
   \label{fig:TGA(epsilon)}
%\end{adjustwidth}
\end{figure}

The parameter $\epsilon$ lets us control the precision of the non-Markovian temporal Gillespie algorithm: the smaller $\epsilon$ is, the more precise the algorithm is, on the other hand, the larger $\epsilon$ is, the faster the algorithm is (Fig~\ref{fig:TGA(epsilon)}).
At $\epsilon=0$, the temporal Gillespie algorithm is maximally accurate, but also slowest, corresponding to the quasi-exact approximation that $\La{t;\Ft}$ stays constant over a single time-step.
Letting $\epsilon\to\infty$ corresponds to the first order cumulant expansion of~\cite{Boguna2014}, and is the fastest, but least accurate. Intermediate $\epsilon$ gives intermediate accuracy and speed, and permits one to obtain the desired accuracy without sacrificing performance. In the case of the SIR process with Weibull-distributed recovery times, $\epsilon=0.1$ gives an error of no more than a few percent (Figs.~\ref{fig:TGA(epsilon)}A--\ref{fig:TGA(epsilon)}D and \ref{fig:validate-nM})---which is usually acceptable as the discrepancy between model and reality can be expected to be larger---with an almost optimal computation time (Figs.~\ref{fig:TGA(epsilon)}E and \ref{fig:speedSIR-nM}).

\section{Discussion}
\label{sec:conclusion}
We have presented a fast temporal Gillespie algorithm for simulating stochastic processes on time-varying networks.
The temporal Gillespie algorithm is up to multiple orders of magnitude faster than current algorithms for simulating stochastic processes on time-varying networks.
For Poisson (constant-rate) processes, where it is stochastically exact, its application is particularly simple.
The algorithm is also applicable to non-Markovian processes, where a control parameter lets one choose the desired accuracy and performance in terms of simulation speed.
We have shown how to apply it to compartmental models of contagion in human contact networks.
The scope of the temporal Gillespie algorithm is more general than this, however, and it may be applied e.g. to diffusion-like processes or systems for which a network description is not appropriate.

\section*{Acknowledgments}
The authors thank Alain Barrat, Rossana Mastrandrea and Julie Fournet for helpful discussions and critical reading of the manuscript
and Thomas L. Vestergaard for help with debugging and code review.
The authors also thank the SocioPatterns collaboration for privileged
access to data sets.
C.L.V. is supported by the EU FET project Multiplex 317532 and M.G. by the French ANR project HarMS-flu (ANR-12-MONU-0018).

% You may title this section "Methods" or "Models".
% "Models" is not a valid title for PLoS ONE authors. However, PLoS ONE
% authors may use "Analysis"
\section*{Methods}
\label{Methods}
The following four subsections contain supporting information to the manuscript:
the first subsection lists notation used in the article (Notation);
the second {details how Monte Carlo simulations were performed (Details on how Monte Carlo simulations were performed)} the third {gives pseudocode for application of the temporal Gillespie algorithm to specific contagion processes on time-varying networks (Algorithms for simulating specific contagion models)}.
Finally, in the fourth subsection we {give pseudocode} for further optimization of the algorithm for empirical networks {by removal of obsolete contacts} (Removing obsolete contacts for an SIR process on empirical networks).

\subsection*{Notation}
\label{app:notation}
\begin{table*}
%\begin{adjustwidth}{-2.25in}{0in}
  \centering
  \caption{{\bf Notation pertaining to the temporal Gillespie algorithm.}}
  \label{tab:notation-TGA}
  \begin{tabular}{|l|l|l|}
    \hline
    {\bf Symbol} & {\bf Description} & {\bf First appearance(s)}\\
    \hline
    $t$ & Real time. & Sec.~\ref{sec:processes}\\
    \hline
    $\Dt$ & Duration of a time-step. & Sec.~\ref{sec:rejectionSampling}\\
    \hline
    $n$ & Time-step number. & Sec.~\ref{sec:temporalGillespie} \\
    \hline
    {$t_n$} & {Time at beginning of time-step $n$: $t_n=n\Dt$.} & {Sec.~\ref{sec:temporalGillespie}} \\
    \hline
    $m$ & Possible transition / transition process. & Sec.~\ref{sec:processes} \\
    \hline
    $\l{m}$ & Transition rate for $m$. & Sec.~\ref{sec:processes} \\
    \hline
    $\Im(t)$ & Function indicating if the transition $m$ may take place at time $t$. & Sec.~\ref{sec:temporalGillespie} \\
    \hline
    $\Oa(t)$ & Set of transition processes at time $t$. & Secs.~\ref{sec:processes},\ref{sec:temporalGillespie} \\
    \hline
    $M(t)$ & Number of transition processes at time $t$. &     Secs.~\ref{sec:processes},\ref{sec:temporalGillespie} \\
    \hline
    $\Oa$ & Set of total possible transitions between two consecutive transition events. &     Secs.~\ref{sec:processes},\ref{sec:temporalGillespie}\\
    \hline
    $M$ & Number of total possible transitions between two consecutive transition events. &     Secs.~\ref{sec:processes},\ref{sec:temporalGillespie}\\
    \hline
    $\Lambda$, $\Lat$ & Cumulative transition rate (at time $t$): $\Lat=\sum_{m\in\Oa(t)}\l{m}$. &     Secs.~\ref{sec:staticGillespie},\ref{sec:temporalGillespie}\\
    \hline
    $\Lt$ & Integrated cumulative transition rate (from $\ts$ to $t$). & Sec.~\ref{sec:temporalGillespie} \\
    \hline
    $\tau$ & Waiting time between two consecutive transitions. & Sec.~\ref{sec:staticGillespie} \\
    \hline
    $S(\tau)$ & Waiting time survival function. & Sec.~\ref{sec:staticGillespie} \\
    \hline
    $\ts$, $\tss$ & Times when the last/next transition took/takes place, respectively. & Sec.~\ref{sec:temporalGillespie} \\
    \hline
    $\ns$, $\nss$ & Time-steps during which the last/next transition took/takes place, respectively. & Sec.~\ref{sec:temporalGillespie} \\
    \hline
    $\tau'$ & Normalized waiting time between two consecutive transition events. & Sec.~\ref{sec:temporalGillespie} \\
    \hline
    $S(\tau')$ & Normalized waiting time survival function. & Sec.~\ref{sec:temporalGillespie} \\
    \hline
    $\tau'\sim\Exp(1)$ & $\tau'$ is exponentially distributed with unit {rate}. & Sec.~\ref{sec:temporalGillespie} (Sec.~\ref{sec:staticGillespie}) \\
    \hline
    {$\Trs$} & {Time per simulation for the rejection sampling algorithm.} & {Sec.~\ref{sec:comparison}} \\
    \hline
    {$\Ttga$} & {Time per simulation for the temporal Gillespie algorithm.} & {Sec.~\ref{sec:comparison}} \\
    \hline
    {$\Ord{x}$} & {Term of order $x$, i.e., $\Ord{x}=a\,x$ for a given constant $a$.} & {Sec.~\ref{sec:comparison}} \\
    \hline
    $\F{m}$ & Filtration for the transition process $m$. & Sec.~\ref{sec:nonMarkovian}\\
    \hline
    $\Ft$ & Union of all $\F{m}$. & Sec.~\ref{sec:nonMarkovian}\\
    \hline
  \end{tabular}
%\end{adjustwidth}
\end{table*}
\begin{table*}
%\begin{adjustwidth}{-2.25in}{0in}
  \centering
  \caption{{\bf Notation pertaining to compartmental contagion models and time-varying networks.}}
  \label{tab:notation-networks}
  \begin{tabular}{|l|l|l|}
    \hline
    {\bf Symbol} & {\bf Description} & {\bf First appearance}\\
    \hline
%    $\Dt$ & Duration of a time-step.\\
    $i$, $j$ & Node. & Sec.~\ref{sec:compartmental}\\
    \hline
    $N$ & Number of nodes in network. & Sec.~\ref{sec:compartmental}\\
    \hline
    $\ct$ & Contact taking place at time $t$ between nodes $i$ and $j$. &
    Sec.~\ref{sec:compartmental}\\
    \hline
    $\kIt$ & Number of infected nodes in contact with $i$ at time $t$ & Sec.~\ref{sec:compartmental}\\
    \hline
    $\kt$ & Average degree (number of contacts per node) of network at time $t$. & Fig.~\ref{fig:speed}\\
    \hline
    $x_i(t)$ & Random variable specifying the state (compartment) of node $i$ at time $t$. & Sec.~\ref{sec:compartmental}\\
    \hline
    $\X{}\in\{\X{1},\X{2}\ldots\X{q}\}$ & Possible node states (compartments). & Sec.~\ref{sec:compartmental}\\
    \hline
    $X_p$ & Number of nodes in state $\X{p}$. & Sec.~\ref{sec:compartmental}\\
    \hline
    $\S$, $\I$, $\R$ & Possible node states in SIS and SIR models of epidemic spreading. & Sec.~\ref{sec:compartmental}\\
    \hline
    $S$, $I$, $R$ & Number of nodes in each of the states $\S$, $\I$, and $\R$, respectively. & Sec.~\ref{sec:compartmental}\\
    \hline
    $\beta$ & Rate of $\S\to\I$ transition of a susceptible node in contact with an infectious node. & Sec.~\ref{sec:compartmental}\\
    \hline
    $\mu$ & Rate of spontaneous $\I\to\R$ or $\I\to\S$ transition of an infectious node. & Sec.~\ref{sec:compartmental}\\
    \hline
    {$\Emean$} & {Mean number of contacts during a single time-step.} & {Sec.~\ref{sec:comparison}} \\
    \hline
    {$\Mmean$} & {Mean number of transition processes per single time-step.} & {Sec.~\ref{sec:comparison}} \\
    \hline
    {$\Qmean$} & {Mean number of transitions that take place per time-step.} & {Sec.~\ref{sec:comparison}} \\
    \hline
    {$\Msi$} & {Mean number of $\S$--$\I$ contacts during a single time-step.} & {Sec.~\ref{sec:comparison}} \\
    \hline
    {$\Imean$} & {Mean number of infectious nodes.} & {Sec.~\ref{sec:comparison}} \\
    \hline
    {$T$} & {Length of a data set describing a time-varying network (in time).} & {Sec.~\ref{sec:comparison}} \\
    \hline
    {$\nsimu$} & {Number of time-steps that are simulated during a single realization.} & {Sec.~\ref{sec:comparison}} \\
    \hline
    {$\Dt_{\rm RS}$} & {Time-step used for rejection sampling when $\l{m}\Dt$ are large, $\Dt_{\rm RS}\leq\Dt$.} & {Sec.~\ref{sec:comparison}} \\
    \hline
    $\mu_0$ & Scale parameter of the Weibull distribution. & Sec.~\ref{sec:nonMarkov-algo}\\
    \hline
    $\gamma$ & Shape parameter of the Weibull distribution. & Sec.~\ref{sec:nonMarkov-algo}\\
    \hline
    $\t{m}$ & Starting time for transition process $m$ (e.g. the time when a node was infected).  & Sec.~\ref{sec:nonMarkov-algo}\\
    \hline
  \end{tabular}
%\end{adjustwidth}
\end{table*}
Tables~\ref{tab:notation-TGA} and \ref{tab:notation-networks} list the notation used in the manuscript. Table~\ref{tab:notation-TGA} gives notation pertaining to the temporal Gillespie algorithm, and Table~\ref{tab:notation-networks}~lists notation pertaining to time-varying networks and compartmental contagion processes.

\subsection*{{Details on how Monte Carlo simulations were performed}}
\label{app:simu}
{All simulations for comparing the speed of algorithms were performed sequentially on a HP EliteBook Folio 9470m with a dual-core (4 threads) Intel Core i7-3687U CPU @ 2.10 GHz. The system had 8 GB 1\,600 MHz DDR3 SDRAM and a 256 GB mSATA Solid State Drive. Code was compiled with TDM GCC 64 bit using} \verb|g++| {with the optimization option} \verb|-O2|{. Speedtests were also performed using} \verb|-O3| {and}  \verb|-Ofast|{, but} \verb|-O2| {gave the fastest results, both for rejection sampling and temporal Gillespie algorithms.}

{For SIR processes simulations were run until $I=0$; for SIS processes simulations were run for $20/(\mu\Dt)$ time-steps (as in Fig.~\ref{fig:validate}) or until $I=0$, whichever happened first.
Between 100 and 10\,000 independent realizations were performed for each data point depending on $\mu\Dt$ (100 for low $\mu\Dt$ and 10\,000 for high $\mu\Dt$).
For simulations on empirical contact data, data sets were looped if necessary.}

\subsection*{Algorithms for simulating specific contagion models}
\label{app:pseudocode}
\renewcommand{\figurename}{Pseudocode}
\setcounter{figure}{0}
We here give pseudocode for the application of the temporal Gillespie algorithm to three specific models:
the first subsection treats the Poissonian SIR process,
the second treats the Poissonian SIS process,
and the third treats a non-Markovian SIR process with recovery times following a general distribution.

We assume in the following that the time-varying network is represented by a list of lists of individual contacts taking place during each time-step. An individual contact, termed \verb|contact|, is represented by a tuple of nodes, \verb|i| and \verb|j|.
The list \verb|contactLists[t]| gives the contacts taking place during a single time-step, \verb|t|, for \verb|t=0,1,...,T_simulation-1|, where \verb|T_simulation| is the desired number of time-steps to simulate.
The state of each node is given by the vector \verb|x|, where the entry \verb|x[i]|$\in\{\verb|S,I,R|\}$ gives the state of node $i$.

As one may always normalize time by the duration of a time-step, $\Dt$, we have in the following set $\Dt=1$. Note that \verb|beta| and \verb|mu| in the code then corresponds to $\beta\Dt$ and $\mu\Dt$, respectively.

\paragraph*{SIR process.}
\label{app:SIR}
The classical SIR model with constant infection and recovery rates is the simplest epidemic model where individuals can gain immunity.
As discussed in the main text, nodes may be in one of three states: susceptible ($\S$), infectious ($\I$), or recovered ($\R$). Two different transition types let the nodes switch state: a spontaneous $\I\to\R$ transition which takes place with rate $\mu$, and a contact-dependent $\S\to\I$ transition which takes place with rate $\beta$ upon contact with an infectious node.
Pseudocode~\ref{fig:codeSIR} shows how the temporal Gillespie algorithm may be implemented for an SIR process on a time-varying contact network.
\paragraph*{SIS process.}
\label{app:SIS}
In the SIS model, nodes can be in one of two states: susceptible ($\S$) or infectious ($\I$). As for the SIR model, two different transition types let the nodes switch state: a spontaneous $\I\to\S$ transition which takes place with rate $\mu$, and a contact-dependent $\S\to\I$ transition which takes place with rate $\beta$ upon contact with an infectious node.

The SIS model is implemented in a manner very similar to the SIR model; an implementation can be found by using the code of Pseudocode~\ref{fig:codeSIR} with lines \verb|07| and \verb|40| removed and line \verb|37| replaced by \verb|x[m] = S|.
C++ code is found at \url{https://github.com/CLVestergaard/TemporalGillespieAlgorithm} for both homogeneous and heterogeneous populations.

\begin{figure*}[p]
%\begin{adjustwidth}{0in}{0in}
\small
\centering
\begin{tikzpicture}
\node at(-4.4,0.18)
{\begin{minipage}[t]{0.5\textwidth}
\begin{Verbatim}[commandchars=\\\{\}]
   \color{green}//Initialize:
{\color{blue}01} {\color{red}FOR} i=1,...,N
{\color{blue}02}   x[i] {\color{red}=} S {\color{green}//set node states to S}
{\color{blue}03} \color{red}ENDFOR
{\color{blue}04} x[root] {\color{red}=} I {\color{green}//set state of root node to I}
{\color{blue}05} m_I {\color{red}=} [root] {\color{green}//list of infected nodes}
{\color{blue}06} N_I {\color{red}=} 1 {\color{green}//number of infected nodes}
{\color{blue}07} N_R {\color{red}=} 0 {\color{green}//number of recovered nodes}
{\color{blue}08} Mu {\color{red}=} mu {\color{green}//cumulative recovery rate}
{\color{blue}09} tau = randexp(1) {\color{green}//draw tau ~ Exp(1)}

   \color{green}//Run through the time-steps:
{\color{blue}10} {\color{red}FOR} t=0,1,...,T_simulation-1
     {\color{green}//Update list of possible S->I transitions:}
{\color{blue}11}   {\color{red}CLEAR} m_SI {\color{green}//S nodes in contact with I nodes}
{\color{blue}12}   {\color{red}FOR} contact {\color{red}in} contactLists[t]
{\color{blue}13}     (i,j) {\color{red}=} contact
{\color{blue}14}     {\color{red}IF} (x[i],x[j]){\color{red}==}(S,I)
{\color{blue}15}       {\color{red}APPEND} i {\color{red}to} m_SI
{\color{blue}16}     {\color{red}ELSE IF} (x[i],x[j]){\color{red}==}(I,S)
{\color{blue}17}       {\color{red}APPEND} j {\color{red}to} m_SI
{\color{blue}18}     {\color{red}ENDIF}
{\color{blue}19}   {\color{red}ENDFOR}
{\color{blue}20}   M_SI {\color{red}= length of} m_si
{\color{blue}21}   Beta {\color{red}=} beta{\color{red}*}M_SI {\color{green}//cumulative infection rate}
{\color{blue}22}   Lambda {\color{red}=} Mu{\color{red}+}Beta {\color{green}//cumulative transition rate}



\end{Verbatim}
\end{minipage}
};

\node at(4.2,0)
{\begin{minipage}[t]{0.5\textwidth}
\begin{Verbatim}[commandchars=\\\{\}]
{\color{green}//Check if a transition takes place:}
{\color{blue}23}   {\color{red}IF} Lambda{\color{red}<}tau {\color{green}//no transition}
{\color{blue}24}     tau {\color{red}-=} Lambda
{\color{blue}25}   {\color{red}ELSE} {\color{green}//at least one transition}
{\color{blue}26}     xi {\color{red}=} 1. {\color{green}//remaining fraction of time-step}
{\color{blue}27}     {\color{red}WHILE} xi{\color{red}*}Lambda{\color{red}>=}tau
{\color{blue}28}       {\color{red}DRAW} z {\color{red}uniformly from} [0,Lambda)
{\color{blue}29}       {\color{red}IF} z{\color{red}<}Beta {\color{green}//S->I transition}
{\color{blue}30}         {\color{red}DRAW} m {\color{red}at random from} m_SI
{\color{blue}31}         x[m] {\color{red}=} I
{\color{blue}32}         {\color{red}APPEND} m {\color{red}to} m_I
{\color{blue}33}         N_I {\color{red}+=} 1
{\color{blue}34}         Mu {\color{red}+=} mu
{\color{blue}35}       {\color{red}ELSE} {\color{green}//I->R transition}
{\color{blue}36}         {\color{red}DRAW} m {\color{red}at random from} m_I
{\color{blue}37}         x[m] {\color{red}=} R
{\color{blue}38}         {\color{red}REMOVE} m {\color{red}from} m_I
{\color{blue}39}         N_I {\color{red}-=} 1
{\color{blue}40}         N_R {\color{red}+=} 1
{\color{blue}41}         Mu {\color{red}-=} mu
{\color{blue}42}       {\color{red}ENDIF}
{\color{blue}43}       xi {\color{red}-=} tau{\color{red}/}Lambda {\color{green}//update remaining fraction}
         {\color{green}//Update list of S->I transitions and rates:}
{\color{blue}44}       {\color{red}REDO} {\color{blue}lines 11-22}
{\color{blue}45}       tau {\color{red}=} randexp(1) {\color{green}//draw new tau}
{\color{blue}46}     {\color{red}ENDWHILE}
{\color{blue}47}   {\color{red}ENDIF}
     {\color{green}//Read out the desired quantities:}
{\color{blue}48}   {\color{red}WRITE} N_I, N_R, ...
{\color{blue}49} {\color{red}ENDFOR}
\end{Verbatim}
\end{minipage}
};
\end{tikzpicture}
  \caption{{\bf Pseudocode for an SIR process with constant and homogeneous transition rates.}
  C++ code for homogeneous and heterogeneous populations is found at \url{https://github.com/CLVestergaard/TemporalGillespieAlgorithm}.}
  \label{fig:codeSIR}
%\end{adjustwidth}
\end{figure*}

\paragraph*{Non-Markovian SIR process.}
\label{app:nM-SIR}
\SaveVerb{muverb}|mu|
\begin{figure*}
\small
%\begin{adjustwidth}{-2.25in}{0in}
\begin{tikzpicture}
\node at(-4.4,0.18)
{\begin{minipage}[t]{0.5\textwidth}
\begin{Verbatim}[commandchars=\\\{\}]
   \color{green}//Initialize:
{\color{blue}01} {\color{red}FOR} i=1,...,N
{\color{blue}02}   x[i] {\color{red}=} S {\color{green}//set nodes states to S}
{\color{blue}03} \color{red}ENDFOR
{\color{blue}04} x[root] {\color{red}=} I {\color{green}//set state of root node to I}
{\color{blue}05} t_inf[root] {\color{red}=} 0 {\color{green}//time of infection = 0}
{\color{blue}06} m_I {\color{red}=} [root] {\color{green}//list of infected nodes}
{\color{blue}07} mus {\color{red}=} [mu(0)] {\color{green}//list of recovery rates}
{\color{blue}08} Mu {\color{red}=} mu(0) {\color{green}//cumulative recovery rate}
{\color{blue}09} N_I {\color{red}=} 1 {\color{green}//number of infected nodes}
{\color{blue}10} N_R {\color{red}=} 0 {\color{green}//number of recovered nodes}
{\color{blue}11} tau = randexp(1) {\color{green}//draw tau ~ Exp(1)}

   \color{green}//Run through the time-steps:
{\color{blue}12} {\color{red}FOR} t=0,1,...,T_simulation-1
     {\color{green}//Update mus if t-t*>=epsilon*mu_avg:}
{\color{blue}13}   {\color{red}IF} t{\color{red}-}t*{\color{red}>=}epsilon{\color{red}*}mu_avg
{\color{blue}14}     {\color{red}CLEAR} mus
{\color{blue}15}     {\color{red}FOR} m {\color{red}in} m_I
{\color{blue}16}       {\color{red}APPEND} mu(t{\color{red}-}t_inf[m]) {\color{red}to} mus
{\color{blue}17}     {\color{red}ENDFOR}
{\color{blue}18}     Mu {\color{red}= sum of} mus
{\color{blue}19}   {\color{red}ENDIF}
     {\color{green}//Update list of possible S->I transitions:}
{\color{blue}20}   {\color{red}CLEAR} m_SI {\color{green}//S nodes in contact with I nodes}
{\color{blue}21}   {\color{red}FOR} contact {\color{red}in} contactLists[t]
{\color{blue}22}     (i,j) {\color{red}=} contact
{\color{blue}23}     {\color{red}IF} (x[i],x[j]){\color{red}==}(S,I)
{\color{blue}24}       {\color{red}APPEND} i {\color{red}to} m_SI
{\color{blue}25}     {\color{red}ELSE IF} (x[i],x[j]){\color{red}==}(I,S)
{\color{blue}26}       {\color{red}APPEND} j {\color{red}to} m_SI
{\color{blue}27}     {\color{red}ENDIF}
{\color{blue}28}   {\color{red}ENDFOR}
{\color{blue}29}   M_SI {\color{red}= length of} m_si
{\color{blue}30}   Beta {\color{red}=} beta{\color{red}*}M_SI {\color{green}//cumulative infection rate}
{\color{blue}31}   Lambda {\color{red}=} Mu{\color{red}+}Beta {\color{green}//cumulative transition rate}
\end{Verbatim}
\end{minipage}
};

\node at(4.2,0)
{\begin{minipage}[t]{0.5\textwidth}
\begin{Verbatim}[commandchars=\\\{\}]
     {\color{green}//Check if transition takes place:}
{\color{blue}32}   {\color{red}IF} Lambda{\color{red}<}tau {\color{green}//no transition}
{\color{blue}33}     tau {\color{red}-=} Lambda
{\color{blue}34}   {\color{red}ELSE} {\color{green}//at least one transition}
{\color{blue}35}     xi {\color{red}=} 1. {\color{green}//remaining fraction of time-step}
{\color{blue}36}     t* {\color{red}=} t {\color{green}//for calculating transition times}
{\color{blue}37}     {\color{red}WHILE} xi{\color{red}*}Lambda{\color{red}>=}tau
{\color{blue}38}       t* {\color{red}+=} tau/Lambda {\color{green}//transition time}
{\color{blue}39}       {\color{red}DRAW} z {\color{red}uniformly from} [0,Lambda)
{\color{blue}40}       {\color{red}IF} z{\color{red}<}Beta {\color{green}//S->I transition}
{\color{blue}41}         {\color{red}DRAW} m {\color{red}at random from} m_SI
{\color{blue}42}         x[m] {\color{red}=} I
{\color{blue}43}         t_inf[m] {\color{red}=} t*
{\color{blue}44}         {\color{red}APPEND} m {\color{red}to} m_I
{\color{blue}45}         N_I {\color{red}+=} 1
{\color{blue}46}       {\color{red}ELSE} {\color{green}//I->R transition}
{\color{blue}47}         {\color{red}DRAW} m {\color{red}from} m_I {\color{red}with weight} mus[m]
{\color{blue}48}         x[m] {\color{red}=} R
{\color{blue}49}         {\color{red}REMOVE} m {\color{red}from} m_I
{\color{blue}50}         N_I {\color{red}-=} 1
{\color{blue}51}         N_R {\color{red}+=} 1
{\color{blue}52}       {\color{red}ENDIF}
{\color{blue}53}       xi {\color{red}-=} tau{\color{red}/}Lambda {\color{green}//update remaining fraction}
         {\color{green}//Update mus:}
{\color{blue}54}       {\color{red}CLEAR} mus
{\color{blue}55}       {\color{red}FOR} m {\color{red}in} m_I
{\color{blue}56}         {\color{red}APPEND} mu(t*{\color{red}-}t_inf[m]) {\color{red}to} mus
{\color{blue}57}       {\color{red}ENDFOR}
{\color{blue}58}       Mu {\color{red}= sum of} mus
         {\color{green}//Update list of S->I transitions and rates:}
{\color{blue}59}       {\color{red}REDO} {\color{blue}lines 20-31}
{\color{blue}60}       tau {\color{red}=} randexp(1) {\color{green}//draw new tau}
{\color{blue}61}     {\color{red}ENDWHILE}
{\color{blue}62}   {\color{red}ENDIF}
     {\color{green}//Read out the desired quantities:}
{\color{blue}63}   {\color{red}WRITE} N_I, N_R, ...
{\color{blue}64} {\color{red}ENDFOR}
\end{Verbatim}
\end{minipage}
};
\end{tikzpicture}
  \caption{{\bf Pseudocode for a non-Markovian SIR process with non-constant recovery rates.}
  The function \protect\UseVerb{muverb} returns the instantaneous recovery rate as function of $(t-\ts)$; for Weibull distributed recovery times, \protect\UseVerb{muverb} is given by Eq.~(\ref{eq:Weibull}).
  C++ code is found at \url{https://github.com/CLVestergaard/TemporalGillespieAlgorithm}.}
  \label{fig:codeSIR-nM}
%\end{adjustwidth}
\end{figure*}

We consider in the main text (Sec.~\ref{sec:nonMarkovian}) an SIR model with non-constant recovery rates; instead of being exponentially distributed (as in the constant-rate SIR model), the individual recovery times, $\tau^{(m)}$, are here Weibull distributed,
\beq
  \tau^{(m)} \sim \gamma\mu_0 \left(\mu_0\tau^{(m)}\right)^{\gamma-1} e^{-\mu_0\tau^{(m)}} \es.
\eeq
As for the classical SIR model, nodes may be in one of three states: susceptible ($\S$), infectious ($\I$), or recovered ($\R$). Two different transition types let the nodes switch state: a contact-dependent $\S\to\I$ transition with constant rate $\beta$ upon contact with an infectious node, and a spontaneous $\I\to\R$ transition which takes place with rate $\mu(t;\t{m})$, given by Eq.~(\ref{eq:Weibull}).

The implementation of the SIR model with non-exponentially distributed waiting times requires some extension of the code for the constant-rate SIR model to account for the heterogeneous and time-dependent recovery rates.
To this end, we introduce the following variables: \verb|t_inf| lists the times at which each node was infected (if applicable); \verb|t*| is the exact time at which the last transition took place; \verb|mu| is a function of time that is called as \verb|mu(t-t_inf[m])| to return the instantaneous recovery rate of \verb|m| at time \verb|t|; \verb|mu_avg| is the expected recovery time of an infected node and is used together with the precision control parameter \verb|epsilon| in the approximate simulation scheme discussed in Sec.~\ref{sec:nonMarkovian}.
Pseudocode~\ref{fig:codeSIR-nM} shows pseudocode for an implementation of such a SIR model with non-constant recovery rates.

\subsection*{{Removing obsolete contacts for an SIR process on empirical networks}}
\label{app:empirical}
{When simulations are carried out on data which are looped due to their finite length, the speed of the temporal Gillespie algorithm may be further increased for processes with an absorbing state, such as the SIR process by removing obsolete contacts to nodes that have entered such a state. Pseudocode~\ref{fig:codeRemoval} shows pseudocode for removing obsolete contacts; its replaces lines 11 to 19 of Pseudocode ~\ref{fig:codeSIR}.}

\begin{figure*}[ht!]
\small
\begin{tikzpicture}
\node at(0,0)
{\begin{minipage}[t]{0.5\textwidth}
\begin{Verbatim}[commandchars=\\\{\}]
{\color{blue}01}   {\color{red}CLEAR} m_SI {\color{green}//S nodes in contact with I nodes}
{\color{blue}02}   {\color{red}FOR} contact {\color{red}in} contactLists[t]
{\color{blue}03}     (i,j) {\color{red}=} contact
{\color{blue}04}     {\color{red}IF} x[i]{\color{red}==}S
{\color{blue}05}       {\color{red}IF} x[j]{\color{red}==}I
{\color{blue}06}         {\color{red}APPEND} i {\color{red}to} m_SI
{\color{blue}07}       {\color{red}ELSE IF} x[j]{\color{red}==}R {\color{green}//remove if x[j]==R}
{\color{blue}08}         {\color{red}REMOVE} contact {\color{red}from} contactLists[t]
{\color{blue}09}       {\color{red}ENDIF}
{\color{blue}10}     {\color{red}ELSE IF} x[i]{\color{red}==}I
{\color{blue}11}       {\color{red}IF} x[j]{\color{red}==}S
{\color{blue}12}         {\color{red}APPEND} j {\color{red}to} m_SI
{\color{blue}13}       {\color{red}ELSE} {\color{green}//remove if (x[i],x[j])==I or x[i]==R}
{\color{blue}14}         {\color{red}REMOVE} contact {\color{red}from} contactLists[t]
{\color{blue}15}       {\color{red}ENDIF}
{\color{blue}16}     {\color{red}ELSE} {\color{green}//remove if x[i]==R}
{\color{blue}17}       {\color{red}REMOVE} contact {\color{red}from} contactLists[t]
{\color{blue}18}     {\color{red}ENDIF}
{\color{blue}19}   {\color{red}ENDFOR}
\end{Verbatim}
\end{minipage}
};
\end{tikzpicture}
  \caption{{\bf Pseudocode for counting possible $\S\to\I$ transitions with removal of outdated contacts.} C++ code is found at \url{https://github.com/CLVestergaard/TemporalGillespieAlgorithm}.}
  \label{fig:codeRemoval}
\end{figure*}

%\section*{References}
% Either type in your references using
% \begin{thebibliography}{}
% Text
% \end{thebibliography}
%
% OR
%
% Compile your BiBTeX database using our plos2015.bst
% style file and paste the contents of your .bbl file
% here.
%
%\bibliography{FastSim}

\begin{thebibliography}{10}
\bibitem{footnote1}
\newblock Note that, while the principal field of application of the algorithm
  is time-varying networks, the algorithm may also be applied to systems where
  a network-based description is not appropriate{, e.g., where transition rates may be modulated by external time-series}.
\bibitem{SecondOrderCumulant}
\newblock A second-order cumulant expansion is proposed in~\cite{Carletti2012}.
  However, while this is generally a better approximation than the first-order
  expansion, it still assumes that $M(t)\gg1$. Furthermore, if the functions
  describing $\l{m}$ are not differentiable, this approach is not applicable.
  This is notably the case for Weibull (or Gamma) distributed waiting times
  with shape parameter $k<2$ as in the example considered here.
\end{thebibliography}

\begin{titlepage}
\clearpage
\setcounter{figure}{0}
\setcounter{page}{20}
\onecolumngrid

~\vspace{3cm}~

\center{\bf\Large Supplementary Figures}
\thispagestyle{empty}

~\vspace{3cm}~
\end{titlepage}

\renewcommand{\figurename}{{\bf Supplementary FIG.}}
\onecolumngrid

\begin{figure*}
    \includegraphics{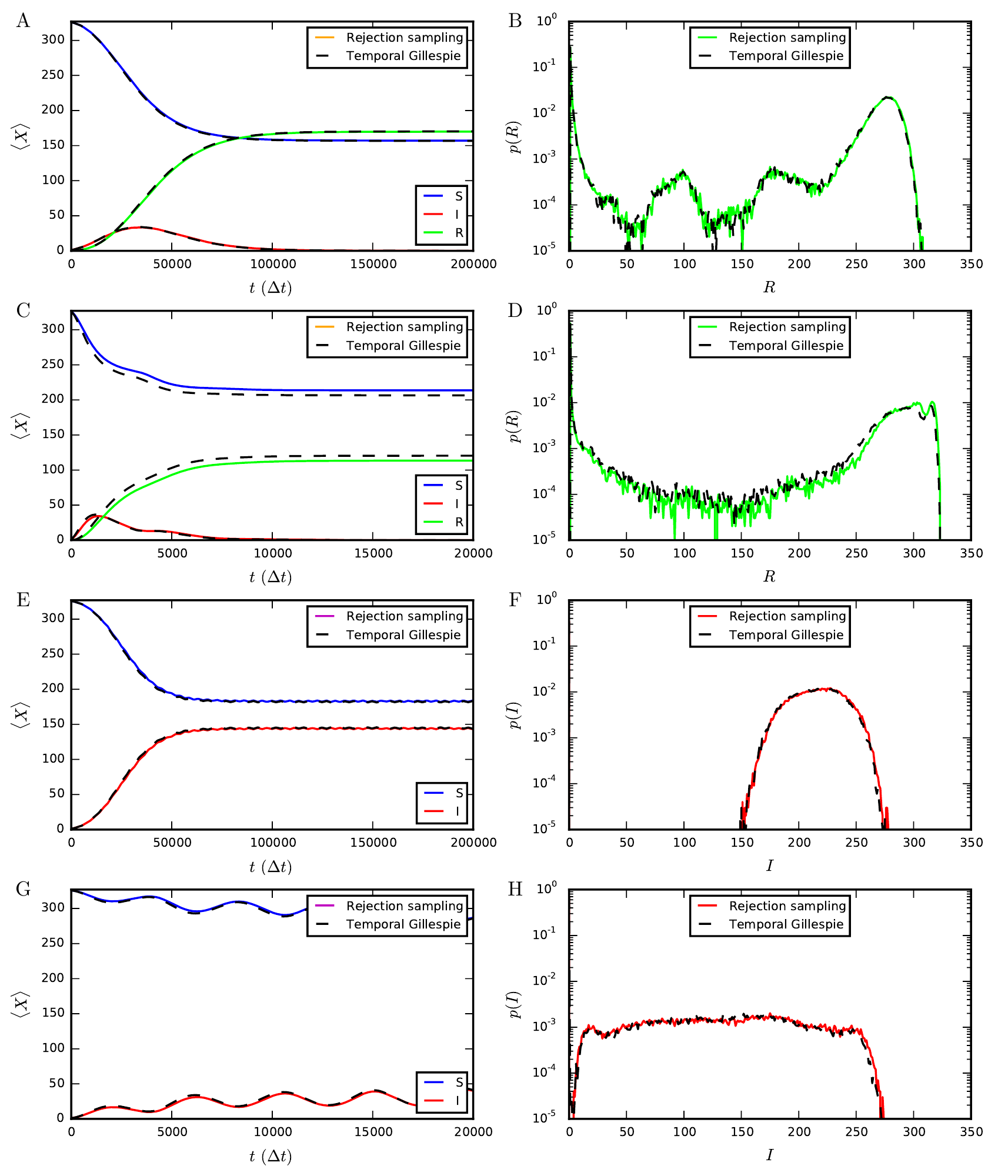}
    \caption{{\bf Numerical results from temporal Gillespie and rejection sampling algorithms for contagion processes taking place on empirical networks.}
    (A)--(D) for a SIR process and (E)--(H) a SIS process.
    (A),(B),(E), and (F) for $\beta\Dt=10^{-2}$ and $\mu\Dt=10^{-4}$; (C),(D),(G), and (H) for $\beta\Dt=10^{-1}$ and $\mu\Dt=10^{-3}$.
    (A),(C) Mean number of nodes in each state of the SIR model as function of time.
    (B),(D) Distribution of final epidemic size (number of recovered nodes when $I=0$) in the SIR model.
    (E),(G) Mean number of nodes in each state of the SIS model as function of time.
    (F),(H) Distribution of the number of infected nodes in the stationary state ($t\to\infty$) of the SIS model.
    All simulations were performed 1\,000\,000 times with the root node chosen at random on a face-to-face contact network recorded in a high school (Table~\ref{tab:empiricalNetworks}).}
    \label{S1_Fig}
\end{figure*}

\begin{figure*}
    \includegraphics{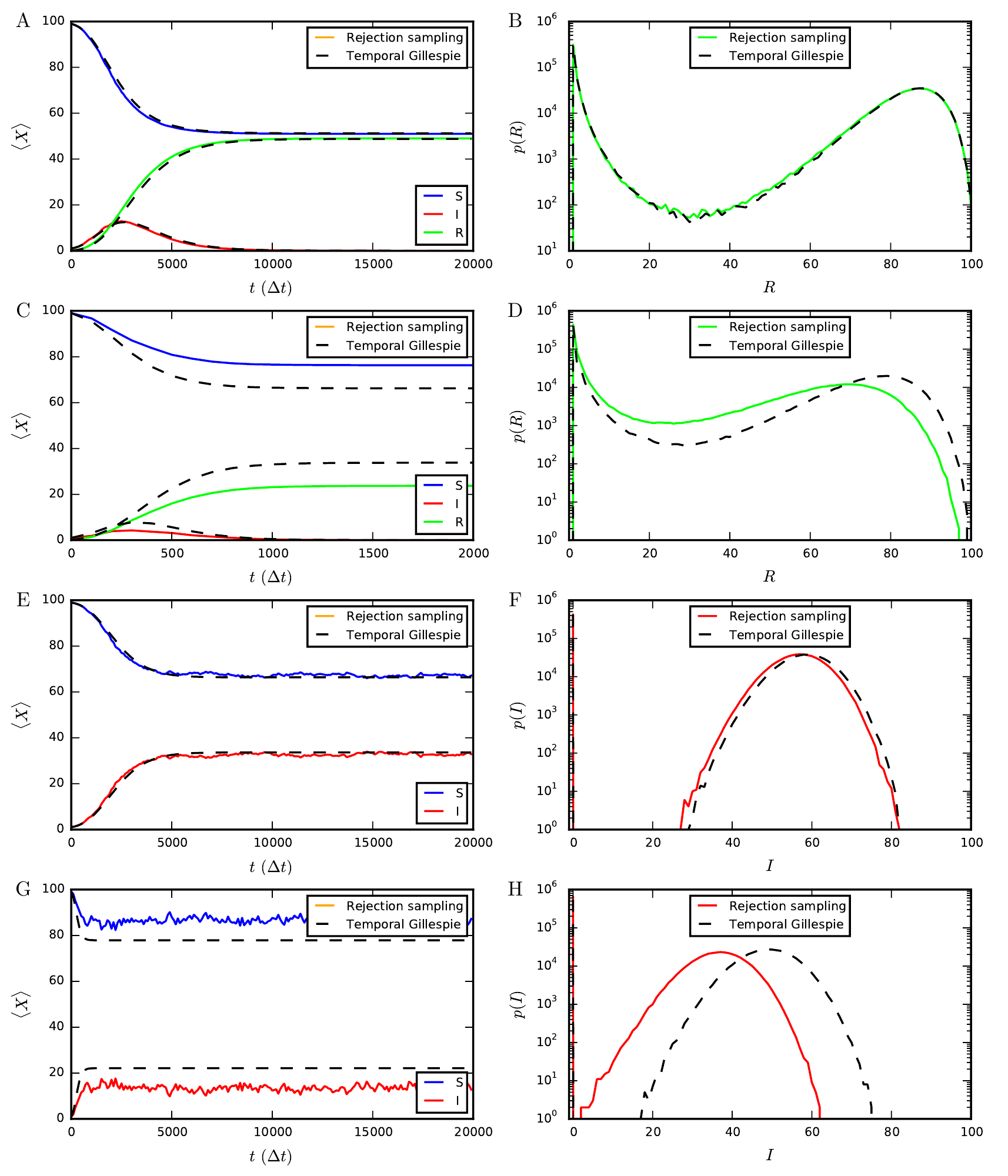}
    \caption{{\bf Comparison of numerical results from temporal Gillespie and rejection sampling algorithms for high transition probability per time-step.} (A)--(D) for a SIR process and (E)--(H) a SIS process.
    (A),(B),(E), and (F) for $\beta\Dt=10^{-1}$ and $\mu\Dt=10^{-3}$; (C),(D),(G), and (H) for $\beta\Dt=1$ and $\mu\Dt=10^{-2}$.
    (A),(C) Mean number of nodes in each state of the SIR model as function of time.
    (B),(D) Distribution of final epidemic size (number of recovered nodes when $I=0$) in the SIR model.
    (E),(G) Mean number of nodes in each state of the SIS model as function of time.
    (F),(H) Distribution of the number of infected nodes in the stationary state ($t\to\infty$) of the SIS model.
    All simulations were performed 1\,000\,000 times with the root node chosen at random on an activity driven network consisting of $N=100$ nodes, with activities $a_i=\eta z_i$, where $\eta=0.1$ and $z_i\sim z_i^{-3.2}$ for $z_i\in[0.03,1)$, and a node formed two contacts when active.}
    \label{S2_Fig}
\end{figure*}

\begin{figure*}
    \includegraphics{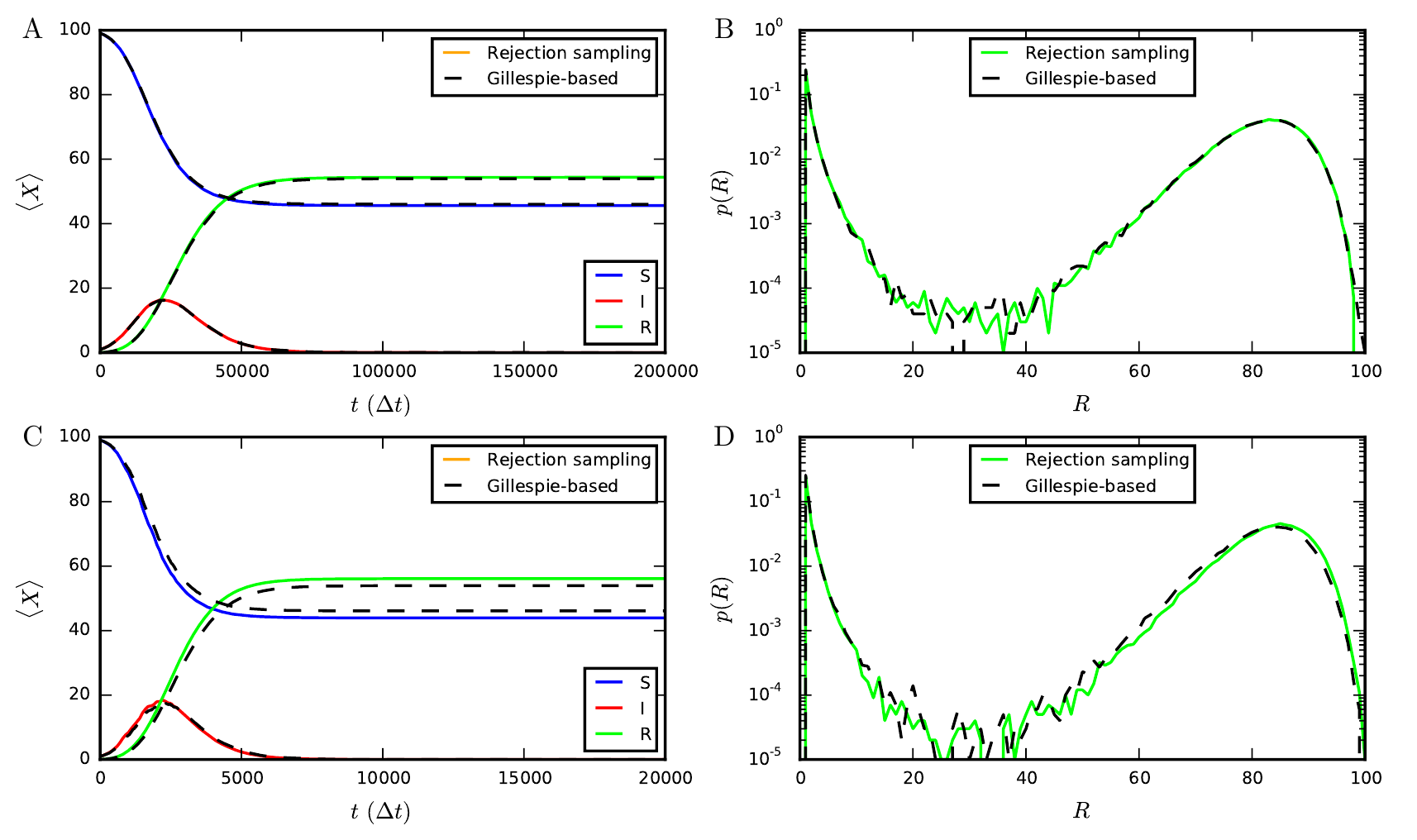}
    \caption{{\bf Comparison of numerical results from temporal Gillespie and rejection sampling algorithms for a non-Markovian SIR process.}
    (a),(c) Mean number of nodes in each state as function of time in the SIR model with Weibull distributed recovery times (Sec.~VII\,A); the parameter controlling the precision of the temporal Gillespie algorithm was set to $\epsilon=0$ (quasi-exact).
    (b),(d) Distribution of final epidemic size (number of recovered nodes when $I=0$).
    (a),(b) $\beta\Dt=10^{-2}$ and $\mu\Dt=10^{-4}$; (c),(d) $\beta\Dt=10^{-1}$ and $\mu\Dt=10^{-3}$. The outcome of the rejection sampling algorithm approaches that of the temporal Gillespie algorithm as $\beta\Dt$ and $\mu\Dt$ become smaller.
    All simulations were performed 100\,000 times with the root node chosen at random on an activity driven network consisting of $N=100$ nodes, with activities $a_i=\eta z_i$, where $\eta=0.1$ and $z_i\sim z_i^{-3.2}$ for $z_i\in[0.03,1)$, and a node formed two contacts when active.
    Nodes' recovery times followed Eq.~(20) with $\gamma=1.5$ and the length of a time-step was $\Dt=1\,{\rm s}$.}
    \label{S3_Fig}
\end{figure*}

\end{document}